\documentclass[12pt]{article}
\usepackage[mathscr]{eucal}
\usepackage{epsfig,amsfonts}
\usepackage[fleqn]{amsmath}
\usepackage{amsthm,amssymb}
\usepackage{graphicx}
\usepackage{hhline}
\usepackage{cite}

\makeatletter
\@addtoreset{equation}{section}
\makeatother

\def\be{\begin{equation}}
\def\ee{\end{equation}}
\def\bdm{\begin{displaymath}}
\def\edm{\end{displaymath}}
\def\bea{\begin{eqnarray}}
\def\eea{\end{eqnarray}}

\def\ri{{\rm i}}

\def\XXint#1#2#3{{\setbox0=\hbox{$#1{#2#3}{\int}$}
    \vcenter{\hbox{$#2#3$}}\kern-.5\wd0}}

\newcommand{\rd}{\mbox{d}}
\newcommand{\re}{\mbox{e}}

\begin{document}

\begin{titlepage}
\begin{flushright}
RUNHETC-2007-03\\
\end{flushright}

\vspace{2cm}

\begin{center}
\begin{LARGE}
 
{\bf Resistively shunted 
Josephson junctions: QFT predictions versus MC results}

\end{LARGE}
\vspace{1.3cm}
                                                                                                   
\begin{large}
                                                                                                   
{\bf  Sergei  L. Lukyanov}$^{1,2}$ {\bf and  Philipp Werner}$^{3}$
                                                                                                   
\end{large}
                                                                                                   
\vspace{1.cm}
                                                                                                   
${}^{1}$NHETC, Department of Physics and Astronomy\\
Rutgers University,
     Piscataway, NJ 08855-0849, USA\\
\vspace{.2cm}
${}^{2}$L.D. Landau Institute for Theoretical Physics\\
  Chernogolovka, 142432, Russia\\

\vspace{.2cm}
and\\
\vspace{.2cm}

${}^{3}$
Department of Physics,
Columbia University\\
538 West, 120th Street,
New York, NY 10027

\vspace{1.0cm}
                                                                                                   
\end{center}
                                                                                                   
\begin{center}

\centerline{\bf Abstract} \vspace{.8cm}
\parbox{11cm}{
During the last fourteen  years several  exact
results were obtained  for the  so-called boundary sine-Gordon model.
In the case of a conformal bulk this 2D boundary quantum field theory  describes  the universal
scaling behavior of the Caldeira-Leggett model of resistively shunted Josephson junctions.
In this work, we use a recently developed
Monte Carlo technique to test some of the analytical predictions.
 }
\end{center}
\vspace{.6cm}
\begin{flushleft}
\rule{3.1 in}{.007 in}\\
{March  2007}
\end{flushleft}
\vfill
\end{titlepage}
\newpage

\tableofcontents

\section{Introduction}

\begin{figure}[ht]
\centering
\includegraphics[width=4cm]{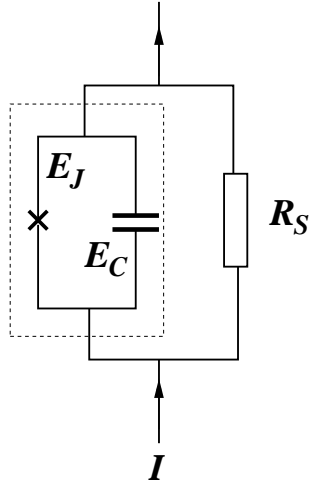}
\caption{Circuit diagram for the shunted
Josephson junction under an external current $I$.
The junction is  characterized by the
Josephson energy $E_J={\hbar \over 2e}\ I_c$, where
$I_c$ is the Josephson critical current,
and
the single
electron charging energy $E_C={e^2\over 2C}$, where $C$  is the  junction
capacitance.
It is shunted by the resistor $R_S\ll R_N$, where $R_N$ is
the normal state tunneling resistance of the
junction.}
\label{fig1}
\end{figure}

For several decades, dissipation
effects in quantum systems have attracted much interest,
because the coupling to the environment affects
the properties of devices displaying macroscopic
quantum coherence and may even induce phase transitions.
These dynamical quantum phase transitions pose challenging theoretical problems 
(see e.g. book \cite{Weiss} and
references therein).
Among the simplest and most fundamental
physical  systems  undergoing a dissipation driven transition
is the resistively shunted
Josephson junction, shown schematically in Fig.\,\ref{fig1}.
\noindent
As long as the Josephson coupling energy
$E_J$ is much larger than the single
electron charging energy $E_C$
a  superconducting state with a well defined
phase difference $\phi$ across the junction is possible.
In this case, the zero-temperature dynamics can be understood
from the  classical equation of  motion for a single particle
in a wash-board potential\ \cite{Tinkham}:
\bea\label{ajjddsj}
{\hbar^2\over 8E_C}\ {\ddot \phi}+
{\partial U_I(\phi)\over\partial \phi}=
-{\hbar
\alpha\over 2\pi}\ {\dot \phi}\ ,
\ \ \ \ \ \ \ \ U_I(\phi)=-E_J\cos(\phi)-{\hbar  I\over 2e}\ \phi\ ,
\eea
where the dimensionless shunt conductance
\bea\label{ssaksalk}
\alpha={\pi \hbar\over  2e^2 }\ R^{-1}_S
\eea
plays the role of the damping strength.
It readily follows from  \eqref{ajjddsj} that
the zero-bias resistance,
$R=
\lim_{I\to 0}{\hbar {\dot\phi} \over 2e I}$,
vanishes (superconducting behavior).
However, for so-called ``ultra small'' junctions characterized
by  $E_J\lesssim E_C$
the phase difference in the superconductors
exhibits large quantum fluctuations and the classical equation \eqref{ajjddsj}
does not provide an adequate description of the system.
Tunneling  between local minima of the
potential energy $U_I(\phi)$\ \eqref{ajjddsj} results in
a finite voltage across the contact (insulating behavior)\
\cite{Zilberman}.
And because the effect of dissipation is to suppress the tunneling rate,
the Superconductor-Insulator (S-I) transition
in the ultra small junction is essentially controlled by the energy dissipation in the shunt.
In the  overdamped limit  $\alpha\to\infty$,
the phase difference $\phi$
localizes in the vicinity of some minimum
of  $U_I(\phi)$ even for $E_C\gg E_J$
and  the device becomes
superconducting\ \cite{ZilbermanA}.
On the other hand, for $\alpha=0$, tunneling events produce
a voltage drop in sufficiently small Josephson junctions.
There have been relatively
little experimental studies of single, ultra small 
Josephson junctions.
The $\alpha-{E_J\over E_C}$ phase diagram
for this important model system
was measured in
Refs.\,\cite{Yagi,Sonin}, and
the experimental results are
consistent with a zero-temperature  S-I transition in the
vicinity of  the line $\alpha=1$.
 
The precise nature of the energy
dissipation in real devices remains unclear.
However, under certain
conditions (see e.g. review \cite{Zaikin}) the S-I transition
is expected to be described by the model introduced by Caldeira and Leggett
in their seminal work on macroscopic quantum tunneling\ \cite{Leggett}.
At zero bias, the Matsubara action of that  model is given by
(here and below $\hbar=k_B=1$)
\bea\label{taction}
{\mathscr   A}&=& \int_{-{\beta\over 2}}^{\beta\over 2}\rd\tau \,
\bigg[\, {(\partial_\tau\phi)^2\over 16 E_C}-E_J\  \cos(\phi)\, \bigg]+
{\alpha \over 8\beta^2 }\
\int^{\beta\over 2}_{-{\beta\over 2}}\rd\tau
\int^{\beta\over 2}_{-{\beta\over 2}}\rd\tau'\ {\big({\phi(\tau)-\phi(\tau')\big)^2}
\over \sin^2({\pi(\tau-\tau')\over \beta})}\ .
\eea
While the first  term in \eqref{taction} readily follows from the classical
equation, the last term introduces dissipation on the quantum level and
results  from integrating out a heat bath characterized by an Ohmic spectral density.
As was argued by Schmid\ \cite{Schmid}, the model \eqref{taction}
exhibits a dissipative phase transition at zero
temperature, which occurs at $\alpha=1$ for sufficiently small values of ${E_J\over E_C}$.
Currently it is believed \cite{Bulgadaev,Guinea,Fisher} that the boundary  of the  transition
lies at   $\alpha=1$ for any value of the ratio
${E_J\over E_C}$, i.e.,
\bea\label{lkssl}
\lim_{\beta\to\infty}\Big({R\over R_S}\Big)=\begin{cases}
0\,,\ \ \ \ \ \ \ \alpha>1\\
1\,, \ \ \ \ \ \ \ \alpha<1
\end{cases}\ ,
\eea
where  $R$ is
the zero-bias resistance.
 
A remarkable feature of the Caldeira-Leggett  model is
that it develops
a universal scaling behavior
characterized by a single energy scale
\bea\label{alsalsal}
E_*\,\propto\, E_C\ \bigg({E_J\over E_C}\bigg)^{\alpha\over \alpha-1}\ ,
\eea
provided
\bea\label{kjssoiuu}
\alpha>1\,, \ \ \ \ \ \  \
\beta E_C\to\infty\, ,\ \ \ \  \ \  \ E_J/E_C\to 0\ .
\eea
In this regime
the model
can be  studied by methods of  Quantum Field Theory (QFT) and
some  physically relevant quantities
can be calculated exactly\ \cite{Zam,Warner, Saleur, BLZ, FLZZ}.

Besides theoretical studies,
numerical simulations of the model\ \eqref{taction}
have been employed to investigate
the location of the S-I phase boundary
at intermediate Josephson coupling \cite{Herrero02} and
the scaling of the resistance with temperature and
dissipation strength \cite{Kimura}.
However, local update Monte Carlo (MC) schemes suffer
from large autocorrelation times at low temperature,
which prevent the simulation in the regime (\ref{kjssoiuu}).
These limitations were overcome with the development of a
cluster  algorithm in Ref.~\cite{WT}, which extends the powerful methods
developed for
spin systems by Swendsen, Wang and Wolff \cite{Swendsen87, Wolff89}
to models described by non-local phase variables.
The algorithm of Ref.~\cite{WT} exploits the symmetry of
the Josephson potential to build the clusters in such a
way that no MC update is rejected and uses an efficient
treatment of the long range couplings in the dissipation term \cite{Luijten95}.
Such non-local updates of the phase configurations dramatically
reduce the autocorrelation time and allow
an ergodic sampling of the discretized version
of action (\ref{taction}), even at low temperatures and strong interactions.
This recent progress in MC techniques
provides  an  opportunity to explore and test the relevance of the
exact scaling results for the quantitative description of the
Caldeira-Leggett model.
The present study is an attempt in this direction. 

The main subject of our interest is
the Matsubara two-point function in the
Caldeira-Leggett model
\bea\label{lsajlsajsa}
G_q(\tau)=\langle\,
\re^{\ri q\phi}(\tau)\ \re^{-\ri q\phi}(0) \, \rangle\, ,
\eea
at $\tau=\beta/2$ and $|q|\leq {1\over 2}$,
in the  experimentally relevant domain of parameters
\bea\label{lasssaslq}
\alpha\geq 1\, ,\ \ \ \ \ \ \ \ \
\beta E_C \gtrsim\, 0.5\times 10^3\, ,\ \ \ \ \ \ \ \ \ \
{E_J\over E_C}\lesssim 0.25\ .
\eea

We begin by describing
$G_q({\beta\over 2})$ at zero temperature  as a function
of the dimensionless parameters $\alpha$, $q$  and  ${E_J\over E_C}$.
This limiting value is expressed in terms of
the Vacuum Expectation Values (VEVs)
\bea\label{lsalssa}
\lim_{\beta\to \infty}G_q
\big({\textstyle{\beta\over 2}}\big)=\langle\, \re^{\ri q\phi}\, \rangle_0^2\,\ .
\eea
In order to clarify  Eq.\,\eqref{lsalssa}
one should note that
in the domain   $\alpha>1$ the global  ${\mathbb Z}$-symmetry,
\bea\label{lskls}
{\mathbb Z}\ :\ \ \phi\to \phi+2\pi n\ \ \ \ \ \  (n=0,\,\pm1,\,\pm 2\ldots)\ ,
\eea
of the theory \eqref{taction}
is spontaneously broken (or equivalently,
the associated quantum dissipative particle is localized at zero temperature).
This implies among other things
that the exponential operators gain nontrivial VEVs
$\langle\, \re^{\ri q\phi}\, \rangle_n$,
where by $\langle\,\cdots\, \rangle_n$ we mean the
expectation value over the ground state in which $\phi$
is localized near $2\pi n$.\footnote{It should be kept in mind  that
$\phi$ in  \eqref{taction} is a non-compact variable $-\infty<\phi<\infty$,
so that the different minima of the  potential $U_0(\phi)=E_J\ \cos(\phi)$
are  distinguishable.}
Notice also
that the  first term in the expansion of
$\langle\, \re^{\ri q\phi}\, \rangle_0$ in $q^2$ gives
the mean phase fluctuation at zero temperature.
In Section\,\ref{Secone}  we combine
an exact nonperturbative
result from  Ref.\,\cite{FLZZ} with  the
conventional perturbative
analyses to make quantitative predictions
for the limiting value\ \eqref{lsalssa}.
Some technical  details of the calculations are presented in the appendix.
For ${E_J\over E_C}=0.125,\, 0.25$,
the  analytical   predictions for \eqref{lsalssa}
($q=0.1,\, 0.25,\, 0.5$) and for  the mean phase fluctuation
are found to be
in good agreement with our MC results for the discretized action
\eqref{taction} (see Figs.\,\ref{fig2} and \ref{fig3}).

Unfortunately the problem of exact calculation of the scaling form of
$G_q({\beta\over 2})$  has not yet been solved.
In Section\,\ref{Sectwo} we calculate the leading
low-temperature correction
to  the   scaling limit of  $G_q({\beta\over 2})$ in the
case  $1<\alpha<2$.
It is proportional to
$(\beta E_*)^{2-2\alpha}$ and
cannot be actually  used without paying attention to higher
order corrections of the form
$\propto (\beta E_*)^{2k(1-\alpha)}$ $(k=2,\,3\ldots)$
(see Fig.\,\ref{fig4}).
In the vicinity  of the phase transition line $\alpha=1$
they can be accounted for within
a Renormalization Group (RG)  treatment.

In Section\,\ref{Sectree} we develop the proper renormalization
scheme based  on a certain choice of the running coupling
constant. Namely,
as it was pointed out in Ref.\,\cite{Kane},
the dimensionless ratio ${R\over R_S}$  of zero-bias resistance
and shunt resistance
(or equivalently,
the dimensionless  mobility of the dissipative particle)  possesses a
universal scaling behavior, i.e., it is a function of the
single scaling variable
$\beta E_* $ in the limit\ \eqref{kjssoiuu}:
\bea\label{lakssoisiul}
{\rho}=\lim_{\beta E_C\to\infty\atop
\beta E_*\,-\,{\rm fixed}}\, \Big({R\over R_S}\Big)\ \ \ \ \ \ \ \ \ \
\ \ \ \ \ \ \  (\alpha>1)\ .
\eea
We shall  argue that the scaling function ${\rho}={\rho}(\beta E_*)$
may serve as  the running coupling (see Fig.\,\ref{fig5}).
This function  was originally
calculated
for
integer values of  $\alpha=2,\,3,\, 4\ldots$ within the Thermodynamic
Bethe Ansatz approach in  Ref.\,\cite{Saleur}.
Later  the closed expression for
${\rho}$ was found for  arbitrary  $\alpha>1$\ \cite{BLZ}.
In  Section\,\ref{subsecFourB}
we confirm the  prediction  from \cite{BLZ} by means of
MC simulations (see Figs.\,\ref{fig6} and \ref{fig7}).
Then, in Section\,\ref{subsecFourC}, we discuss
the  anomalous dimensions of exponential operators $\re^{\ri q\phi}$
and perform the  RG re-summation of  perturbative expansions for
the two-point function
$G_q({\beta\over 2})$.
The obtained
results for  $G_q({\beta\over 2})$ in the vicinity
of the S-I phase transition line $\alpha=1$
have been confirmed by the MC simulations (see Fig.\,\ref{fig8}).
 
A recent numerical study \cite{WT} found
that the S-I phase boundary in the Caldeira-Leggett  model
is a line of fixed
points with continuously varying exponents.
The universal behavior in this case
is  governed  by  a certain  one-parameter family
of  boundary Conformal Field
Theories (CFTs) whose
modulus
can be labeled by the limiting value
\bea\label{ajsksajas}
{\rho}=\lim_{\beta E_C\to \infty\atop
E_J/E_C\,-\,{\rm fixed}}
\Big({R\over R_S}\Big)\ :\ \ \ \ \ \ \ \ \ \ \ \ 0< \rho< 1\ \ \ \ \ \
\  \ \ \  \ \ \ (\alpha=1)\ .
\eea
Nowadays  this boundary  CFT family   is relatively well understood
due to
the works \cite{Klebanov, Maldacena, Polchinski,Thorlacius}  which are
mainly motivated by  problems from   String Theory (see e.g. \cite{Sen}).
In particular the exact value of the critical exponent $\eta_q$
describing the universal
behavior of the two-point correlation function
\eqref{lsajlsajsa},
\bea\label{aslsasa}\label{laksla}
G_q(\tau)\propto |\tau|^{-2\eta_q}\  \ \ \ \  \ \ \
(\, E_C^{-1}\ll \tau\ll \beta\, )\  ,
\eea
can be extracted directly from results of Ref.\,\cite{Polchinski}
and reads explicitly
as
\bea\label{ssalkiuy}
\eta_q={\textstyle{1\over\pi^2}}\ \arcsin^2\big(
\sqrt{\rho}\,\sin(\pi q)\big)\ \ \ \ \ \ \ \
\ \ \  \ \ \big(\, |q|\leq {\textstyle{1\over 2}}\, \big)\ .
\eea
In Section \ref{Sectfour} we briefly
recall some facts concerning the boundary CFTs associated
with the critical line  $\alpha=1$ and check the analytical expression for $\eta_q$
by means of MC simulations.
Our numerical results confirm the  prediction\ \eqref{ssalkiuy}
for $q=0.25$ and $0.5$   (see Fig.\,\ref{fig9}).
 
Finally, in Section\,\ref{Sectfive},
we briefly discuss the applicability 
of the scaling results in a wider domain
of parameters than \eqref{lasssaslq}.
As an illustration, we study the dependence
$\rho=\rho({E_J\over E_C})$ \eqref{ajsksajas}
along the critical line $\alpha=1$
for  all values  ${E_J\over E_C}\geq 0$
(see Fig.\,\ref{fig11}).

\section{Vacuum expectation values of  exponential operators \label{Secone}}

\subsection{Small -- ${E_J\over E_C}$ expansion of $\langle\, 
\re^{\ri q\phi}\, \rangle_0$}

It was realized  a while ago \cite{Calan} that   the 
Caldeira-Leggett model in the scaling regime\ \eqref{kjssoiuu}
can be thought of as an effective theory
describing the dynamics of boundary degrees of 
freedom for the Boundary Sine-Gordon (BSG) model defined by the Euclidean  action
\bea\label{lksslasas}
{\mathscr   A}_{\rm BSG}= {1\over 4\pi}
 \int_{-\infty}^0\rd x\int_{-{\beta\over 2}}^{\beta\over 2}
\rd\tau
 \Big[ (\partial_x
{\boldsymbol \varphi})^2+
 (\partial_\tau {\boldsymbol \varphi})^2 \Big]-
2\mu_B \int_{-{\beta\over 2}}^{\beta\over 2}
 \rd \tau\, \cos(\sqrt{g} {\boldsymbol \varphi}_B)
\ .
\eea
Here
${\boldsymbol \varphi}=
{\boldsymbol \varphi}(x,\tau)$ is a one component
uncompactified Bose field,
$(x,\,\tau)$ are coordinates on the Euclidean half-cylinder, $x\leq 0$,\ $\tau\equiv
\tau+\beta$, and
\bea\label{kssksa}
{\boldsymbol  \varphi}_B(\tau)\equiv {\boldsymbol \varphi}
(0,\tau)\ .
\eea
The interaction in \eqref{lksslasas} acts only on 
the boundary $x=0$ and is controlled by the two parameters $g$ and $\mu_B$.
{}For $0<g<1$ the model is a CFT -- a free Bose field with
free boundary condition at $x=0$ -- perturbed by the relevant
boundary operator
$2 \cos(\sqrt{g} {\boldsymbol \varphi}_B)$ of dimension $g$.
We assume that this boundary 
operator is normalized such that the two-point correlation function takes the following asymptotic form 
\bea\label{lsasalsa}
\big\langle\, 2
\cos(\sqrt{g} {\boldsymbol \varphi}_B)(\tau)\ 
2\cos(\sqrt{g} {\boldsymbol \varphi}_B)(\tau')\, \big\rangle\to
2\,  |\tau-\tau'|^{-2g}\ \ \ \ {\rm as}\ \ |\tau-\tau'|\to 0\ .
\eea
Under this 
normalization the parameter $\mu_B$ has the dimension $[\,energy\,]^{1-g}$.

The original Callan-Thorlacius
observation \cite{Calan} is that
upon fixing the boundary values of ${  \varphi}_B$ and 
integrating out the 
bulk part of the field  ${  \varphi}$,
the  action \eqref{lksslasas} reduces to
the effective
action \eqref{taction} without the Coulomb  term provided
\bea\label{lkss}
g=\alpha^{-1}\ ,
\eea
and
the variable $\phi$ is replaced 
by  $\sqrt{g}\varphi_B$.
A superficial examination reveals
that the model \eqref{taction} without the Coulomb  term
is perturbatively ill defined.  It  
 has to be equipped with an ultraviolet (UV) cutoff and needs renormalization.
As long as the  energy scale  $E_C$ remains finite but
essentially exceeds  all other energy scales, the  
Coulomb term in \eqref{taction}
just provides the desired regularization with the
cutoff energy $\Lambda\propto  \alpha E_C$.
In what follows it will be convenient for us to specify the cutoff energy scale
unambiguously, by 
\bea\label{kjsshkjsa}
\Lambda={4\re^{\gamma_E}\over \pi}\ \alpha E_C=2.267\ldots\times  \alpha E_C\ ,
\eea
where $\gamma_E=0.5772\ldots$ is Euler's constant. 
The dimensionless parameter 
\bea\label{lslsaas}
\epsilon={E_J\over 2\Lambda}=0.220\ldots \times \alpha^{-1}\ {E_J\over E_C}\ ,
\eea
in turn
will be treated as the
bare coupling.
Consistent removal of the UV divergence requires that the bare coupling
be given a dependence on the cutoff scale
such that the
limit $\lim_{\Lambda\to\infty}\big(\Lambda^{1-g}\, \epsilon(\Lambda)\big)$
exists  and coincides
with the dimensionful parameter of the 
renormalized action\ \eqref{lksslasas}.
One can always choose   
$\epsilon(\Lambda)$
to satisfy the simple RG flow equation
\bea\label{alskask}
-{\Lambda}\, {\rd\epsilon\over \rd\Lambda}=(1-g)\ \epsilon\ ,
\eea
i.e.,  
\bea\label{jssiusia}
\mu_B=\epsilon\, \Lambda^{1-g}\ .
\eea

The above 
relation 
immediately implies that
Matsubara
operators  
in the dissipative quantum mechanics~\eqref{taction}
can be thought of as bare boundary   
fields. Hence, they 
should admit  asymptotic expansions in terms of
the renormalized  boundary  fields from QFT \eqref{lksslasas}.
In  what follows  we consider
such an expansion for 
the exponential operators.

The simplest and  most fundamental boundary fields  in QFT \eqref{lksslasas}
are the exponential fields
\bea\label{kshsak}
{\boldsymbol {\cal O}}_q=
\re^{\ri q\sqrt{g}
{\boldsymbol \varphi}_B}
\ ,\ \ \ \ \ \ \ {\boldsymbol {\cal O}}^{\dagger}_q=
 {\boldsymbol {\cal O}}_{-q}\ ,
\eea
transforming 
irreducibly  under  the symmetry \eqref{lskls}:
\bea\label{laslsahl} 
{\mathbb Z}\ :\ \ 
{\boldsymbol {\cal O}}_q\to \re^{2\pi\ri q n}\, {\boldsymbol {\cal O}}_q\ .
\eea
Below, the  field ${\boldsymbol {\cal O}}_q$ is always understood as a {\it renormalized}
boundary field
of  definite scaling 
dimension 
\bea\label{klaksa}
d_{q}=g\, q^2\ ,
\eea
 subjected  to the UV renormalization condition
\bea\label{lksasaha}
\langle\, {\boldsymbol {\cal O}}_q(\tau)
{\boldsymbol {\cal O}}^{\dagger}_q(0)\, \rangle\to |\tau|^{-2 d_q}\ 
\ \ \ \ {\rm as}\ \ \ \ |\tau|\to 0\ .
\eea
The  general  structure of the asymptotic  expansion of the
bare   operator $\re^{\ri q\phi}$ readily
follows from the standard RG arguments and
from the global 
${\mathbb Z}$-symmetry \eqref{lskls},\,\eqref{laslsahl}\footnote{Although
the ${\mathbb Z}$-symmetry
is spontaneously broken in the domain $\alpha>1$,
the operator valued relations
still possess this invariance.}${}^{,}$\footnote{Here and
below we use the symbol $\asymp$ to emphasize that the
relation holds in a formal  asymptotic sense without
any  reference to convergence issues.}:
\bea\label{lasajlsa}
\re^{\ri q\phi}(\tau)\,\asymp \sum_{k=0,\pm 1,\pm 2\ldots}
C^{(k)}(q,\epsilon)\ \Lambda^{-d_{q-k}}\ {\boldsymbol {\cal O}}_{q-k}(\tau)+\ldots\ .
\eea
Here, 
the explicit dependence on the cutoff scale $\Lambda$ is used
to show the relative smallness of various terms and
the dots  in \eqref{lasajlsa} stand for the renormalized
boundary fields
 of the form $\partial_\tau {\boldsymbol {\cal O}}_{q-k},\, 
\partial^2_\tau {\boldsymbol {\cal O}}_{q-k},\,
(\partial_{\tau}{\boldsymbol \varphi}_B)^2{\boldsymbol {\cal O}}_{q-k}\ldots$
whose scaling 
dimensions are given by  $d_{q-k}+N$  with  $N=1,\,2\ldots$\ .
An obvious consequence of the $\phi\to-\phi$ invariance
of the action \eqref{lksslasas} is that
the  renormalization constants $C^{(k)}(q,\epsilon)$ in \eqref{lasajlsa} satisfy the relation
\bea\label{aslassal}
C^{(k)}(q,\epsilon)=C^{(-k)}(-q,\epsilon)\ .
\eea
The  renormalization constants do not carry any  information on the 
infrared environment
and therefore  admit  regular  power series expansions in the bare coupling:
\bea\label{slsaskjsa}
 C^{(k)}(q,\epsilon)=
\epsilon^{|k|}\ \sum_{j=0}^\infty C_j^{(k)}(q)\ \epsilon^{2j}\ .
\eea
The  expansion coefficients are in principle computable
within perturbation theory.
In particular, without tedious calculations, one finds
\bea\label{slsaskjsaiu}
C_0^{(0)}=1\ ,
\eea
provided the cutoff scale is
defined  by  Eq.\,\eqref{kjsshkjsa}.
The explicit expression for $C_{0}^{( 1)}(q)=C_{0}^{( -1)}(-q)$ and 
 $C_1^{(0)}(q)$ 
is  somewhat cumbersome  
and  relegated to the appendix (see Eqs.\,\eqref{aslasksala} and \eqref{askjsa}).

Contrary to the  renormalization constants, 
VEVs of  the boundary operators  are  important
characteristics of the infrared environment. They
do not  possess a regular expansion in 
powers of $\epsilon$
and, hence, can not
be calculated perturbatively.
In Ref.\cite{FLZZ} the following 
explicit form for the VEVs of  renormalized 
exponential fields\ \eqref{kshsak}
was proposed:
\bea\label{uysia}
\langle\,  {\boldsymbol {\cal O}}_q\,\rangle_{n}=\re^{2\pi\ri q n}\
\big(\mu_B\big)^{d_q\over 1-g}\
D(q)\ ,
\eea
where the function $D(q)$ for $| \Re e\, q|\leq {1\over 2 g}$ is given by
the integral
\bea\label{laljajsy}
D(q)&=&\bigg[{2^g\pi\over \Gamma(g)}\bigg]^{g q^2\over 1-g}\
\exp\bigg\{
\int_0^{\infty}{\rd t\over t}\times \\ && \bigg[{\sinh^2(gqt) 
(\re^{t}-1+\re^{t(1-g)}+\re^{-g t})
\over 2\sinh(g t)\sinh(t) \sinh((1-g) t)}-gq^2 \, \Big(
{1\over \sinh((1-g) t)}+\re^{-t}\Big)\, \bigg]\bigg\}\, ,\nonumber
\eea
and is defined through analytic continuation outside this domain.

Combining  now the exact result\,\eqref{uysia} where $\mu_B$ is given 
by \eqref{jssiusia} with 
the asymptotic  expansion \eqref{lasajlsa}, one 
obtains:
\bea\label{lasa}
\langle\,\re^{\ri q\phi}\rangle_0\,= \sum_{k=0,\pm 1,\pm 2\ldots}
C^{(k)}(q,\epsilon)\, D(q-k)\ \, \epsilon^{d_{q-k}\over 1-g}\, \Big[\,1+
O\big(\epsilon^{{\rm min}(1+2g,2)\over 1-g}\big)\, \Big]\ .
\eea
The remaining term $O(\ldots)$
in \eqref{lasa} requires some explanations.
Namely, to transform
the operator expansion  \eqref{lasajlsa}
into the VEV expansion,
small-$\epsilon$  corrections to   the BSG  vacuum
state should be   taken  into account.
One can show that 
at the leading order this gives rise   to the  term 
$\propto \epsilon^{{d_{q-k}+1+2g\over 1-g}}$
in \eqref{lasa}.
The correction $\propto \epsilon^{d_{q-k}+2\over 1-g}$, in turn,  is 
a contribution of  the  boundary field
$(\partial_{\tau}{\boldsymbol \varphi}_B)^2{\boldsymbol {\cal O}}_{q-k}$ of
scaling 
dimension $d_{q-k}+2 $ whose 
VEV is currently not known.

\subsection{Comparison with MC results \label{SeconeA}}
 
For the numerical evaluation
of $\langle\, \re^{\ri q\phi}\, \rangle_0$ with $|q|\leq {1\over 2}$
we  included only the contribution of  the renormalized fields
${\boldsymbol {\cal O}}_{q}$ and 
${\boldsymbol {\cal O}}_{q\pm 1}$ 
in the asymptotic expansion\ \eqref{lasa}.
Furthermore, the perturbative
series for the renormalization constants $C^{(\pm 1 )}(q,\epsilon)$ and
$C^{(0)}(q,\epsilon)$  were  
truncated   at the leading  order $(\propto \epsilon$)
and   at the  second-order $(\propto \epsilon^2$), respectively.
Thus our  approximation reads  as follows:
\bea\label{slsasalk}
&&\langle\,\re^{\ri q\phi}\rangle_0=
\epsilon^{g q^2\over 1-g}\  \Big[\,  D(q) +C^{(0)}_1(q) D(q)\ \epsilon^2+
C^{(1)}_0(q)\ D(1-q)\ \epsilon^{1+{g(1-2 q)\over 1-g}}+\nonumber\\ &&\ \ \ \
C^{(1)}_0(-q)\ D(1+q)\ \epsilon^{1+{g(1+2 q)\over 1-g}} + O\Big(\,\epsilon^{{\min }(4,\,
{1+2g\over 1-g},\, 3+{ g(1-2|q|)\over 1-g})}\, \Big)\, \Big]\, \  \ \ \big(\,|q|\leq {\textstyle{1\over 2}}\,\big)\ .
\eea
Explicit formulas for the coefficients $C^{(0)}_1(q)$ and $C^{(1)}_0(q)$
can be found  in the appendix.
It deserves to be mentioned that
for ${\rm min}\big({3\over 4+2 q},{1\over 1+2 q}\big)\leq g<1$ and  $0\leq q \leq{1\over 2}$ the
last explicit  term in 
expansion \eqref{slsasalk}  does not   exceed 
the reminder  $ O(\ldots)$ and therefore should
be ignored.\footnote{ 
Notice that  it  has  poles  in this domain of  parameters.
In order to clarify  the general  source of the
singularities  of the function $D(q)$\ \eqref{laljajsy},
the definition of  renormalized fields 
\eqref{kshsak}  should be critically reexamined.
${\boldsymbol{\cal O}}_q$ have been  described  as boundary fields
with the  definite scaling dimensions\ \eqref{klaksa} and
${\mathbb Z}$-charges \eqref{laslsahl}
satisfying the renormalization condition\ \eqref{lksasaha}. In general
these requirements are sufficient to specify ${\boldsymbol{\cal O}}_q$,
except when certain
resonance conditions are satisfied.
We say that the renormalized  field ${\boldsymbol{\cal O}}^{(1)}$ has a 
$n$-th order resonance with the field ${\boldsymbol{\cal O}}^{(2)}$
if the scaling  dimensions of these fields satisfy the equation
$d^{(1)}=d^{(2)}+n(1-g)$,
with some non-negative integer $n$. If this resonance condition is satisfied
an obvious ambiguity,
${\boldsymbol{\cal O}}^{(1)}\to {\boldsymbol{\cal O}}^{(1)}+const\ (\mu_B)^n\
{\boldsymbol{\cal O}}^{(2)}$,
in  defining the renormalized  field ${\boldsymbol{\cal O}}^{(1)}$
with  the scaling dimension $d^{(1)}$ typically results in the
logarithmic scaling of ${\boldsymbol{\cal O}}^{(1)}$.
For example  the simple  pole of  the function  $D(q)$ at $q={1\over 2g}$
just indicates that the field ${\boldsymbol {\cal O}}_{{1\over 2g}}$
has a
first order resonance with ${\boldsymbol {\cal O}}_{{1\over 2g}-1}$.}
For the same reason the  third term in \eqref{slsasalk}   should  be neglected 
if ${3\over 4-2 q}\leq g<1$ and  $0\leq q \leq{1\over 2}$.
Let us also note that  the 
first nontrivial term of 
the expansion of $\langle\,\re^{\ri q\phi}\rangle_0$ in a 
power series of $q^2$ is simply related to the mean phase fluctuation 
\bea\label{ksskas}
\langle\,\re^{\ri q\phi}\,\rangle_0=1-\textstyle{q^2\over 2}\ \langle\,(\phi-{\bar\phi})^2\,\rangle_0+O(q^4)\ ,
\eea
where ${\bar\phi}=\langle\,\phi\,\rangle_0=0$. Therefore Eqs.\,\eqref{slsasalk} and \eqref{ksskas}
allow one to calculate  $\langle\,(\phi-{\bar\phi})^2\,\rangle_0$ at zero temperature
for sufficiently small values of the  ratio
${E_J\over E_C}$.

The analytical prediction \eqref{slsasalk} has been tested by means 
of MC simulations 
performed 
on the discretized action
\bea\label{MCtaction}
{\mathscr   A_\text{MC}}= -\sum_{i=0}^{N-1} 
\bigg[\, {\cos(\phi_{i+1}-\phi_i)\over 8 \Delta \tau E_C}+\Delta\tau E_J\  \cos(\phi_i)\, \bigg]+
{\alpha \over 4N^2 }\ 
\sum_{i<j} {\big({\phi_i-\phi_j\big)^2}
\over \sin^2({\pi(i-j)\over N})}\ ,
\eea
where $N$ 
is the number of time-slices of size 
$\Delta\tau=\beta/N$. The algorithm of Ref.~\cite{WT} combines local updates in the Fourier components 
\bea\label{fourier}
\tilde \phi_k = \sum_{n=0}^{N-1}\re^{2\pi{\rm i} nk/N}\, \phi_n
\label{fouriercoefficient}
\eea
(required for ergodicity) with rejection-free cluster updates which can shift entire 
segments of phase variables from one 
minimum of the cosine-potential to another.

\begin{figure}[t]
\centering
\includegraphics[width=10cm, angle=-90]{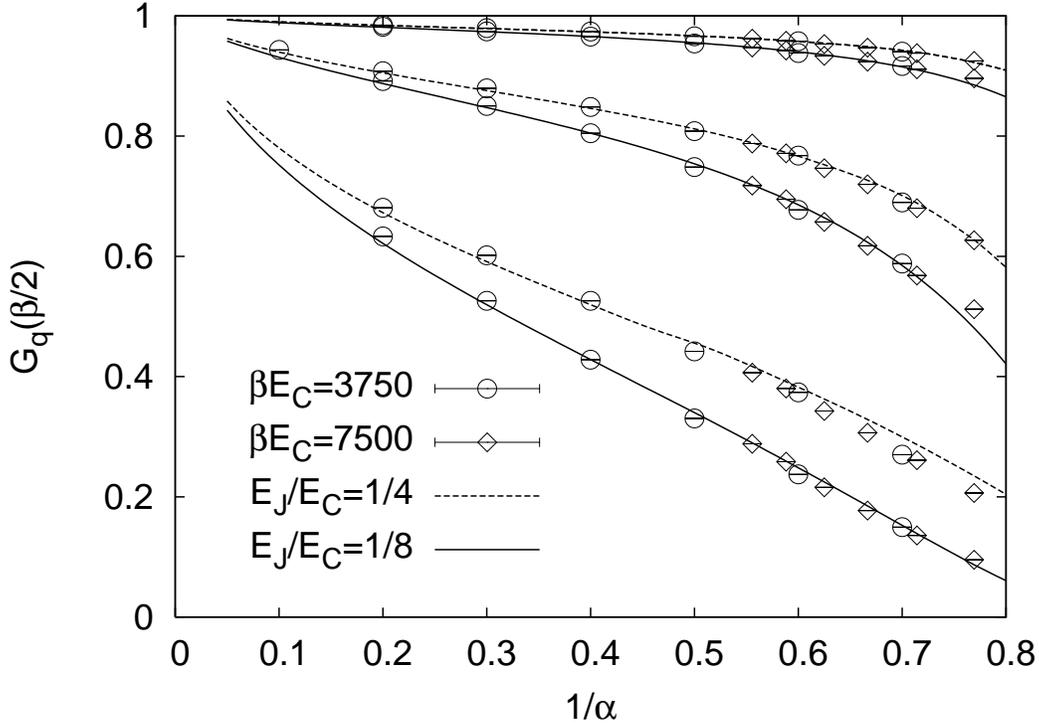}
\caption{ $G_q({\beta\over 2})$ as a function of $\alpha^{-1}$
for  $q=0.1,\, 0.25,\, 0.5$  (from top to bottom). 
The MC data were obtained for ${E_J\over E_C}={1\over 4}, {1\over 8}$  with lattice spacing $\Delta\tau E_C={1\over 4}$.
The circles and diamonds correspond to
$\beta E_C=3750$ and $\beta E_C=7500$, respectively.
The  dashed and solid lines
are the theoretical predictions \eqref{slsasalk} for the VEVs $\langle\,\re^{\ri q\phi}\rangle_0^2$
for  ${E_J\over E_C}={1\over 4}$ and
${E_J\over E_C}={1\over 8}$, respectively.}
\label{fig2}
\end{figure}

\begin{figure}[ht]
\centering
\includegraphics[width=10cm, angle=-90]{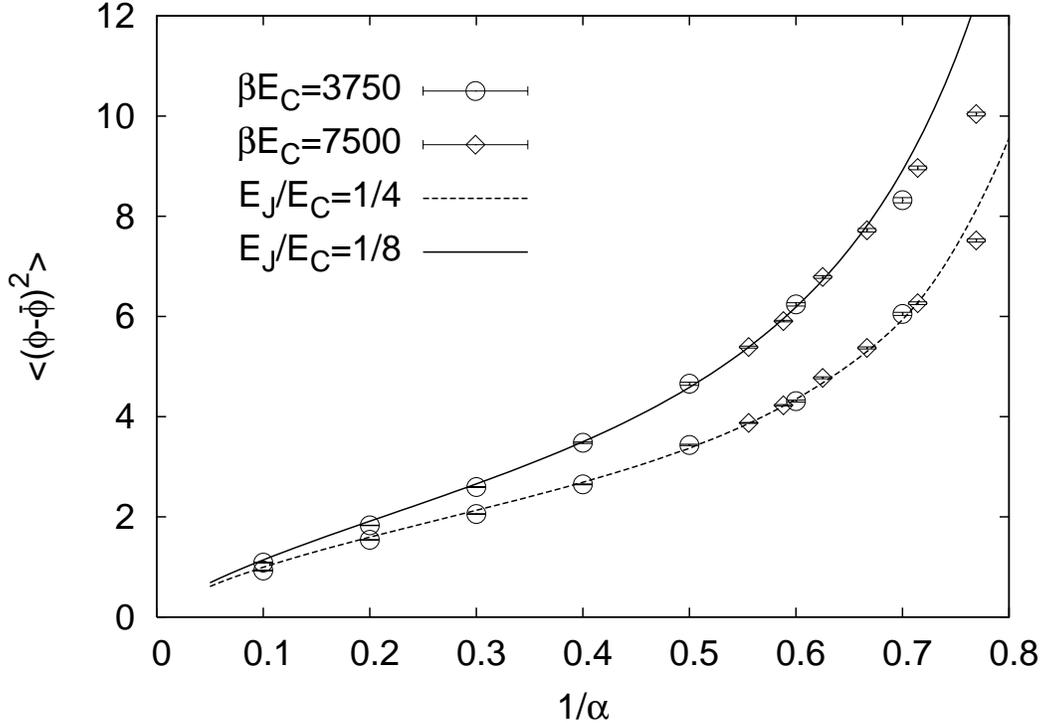}
\caption{
The mean phase fluctuation
as a function of $\alpha^{-1}$.
The MC data were obtained  for ${E_J\over E_C}={1\over 4}, {1\over 8}$  with
lattice spacing $\Delta\tau E_C={1\over 4}$.
The circles and diamonds correspond to
$\beta E_C=3750$ and $\beta E_C=7500$, respectively.
The  dashed and solid lines
are theoretical predictions for the VEV 
$\langle\, (\phi-{\bar\phi})^2\,\rangle_0$ with 
${E_J\over E_C}={1\over 4}$ and
${E_J\over E_C}={1\over 8}$, respectively.} 
\label{fig3}
\end{figure}

The expectation value of $(\phi-\bar\phi)^2$ can 
be obtained directly from the time domain variables $\phi_n$.
In order to measure the correlation functions (\ref{lsajlsajsa}) efficiently, it is useful to note that 
\bea
G_q(n\Delta\tau) = 
\frac{1}{N}\sum_{k=0}^{N-1}\re^{-2\pi{\rm i}  nk/N}\ 
\langle\, |\tilde \Phi_{q,k}|^2\, \rangle\ , 
\eea
where ${\tilde \Phi_{q,k}}$ is the $k$-th component of the Fourier 
transform of $\re^{{\rm i} q\phi}$. 
By computing 
$\tilde \Phi_{q,k}$ (in a time $O(N \log N)$ 
using fast Fourier transformation) and sampling  
$|\tilde \Phi_{q,k}|^2$ in the MC simulation, one can therefore harness all the information contained in the phase configuration without considering each pair of phase variables. 

Results of   the numerical work are depicted in 
Figs.\,\ref{fig2} and \ref{fig3}. 
Similar results were found for ${E_J\over E_C}=
{1\over 16}$ and ${3\over 16}$. 
No essential dependence of the simulation results
on the discretization parameter 
$\Delta\tau  E_C$ was observed for $\Delta\tau E_C\lesssim 0.5$.

\section{Leading  low-temperature 
correction for $G_q({\beta\over 2})$
\label{Sectwo}}

\begin{figure}[t]
\centering
\includegraphics[width=10cm, angle=-90]{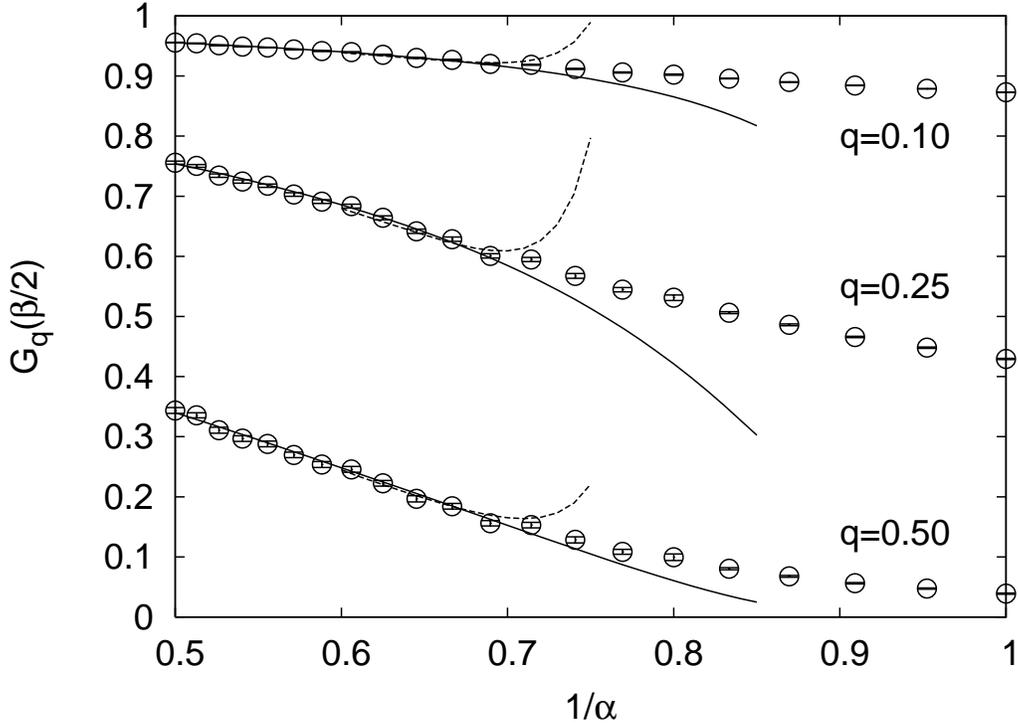}
\caption{
$G_q({\beta\over 2})$ as a function of $\alpha^{-1}$
for  $q=0.1,\, 0.25,\, 0.5$.
The MC data were obtained  for ${E_J\over E_C}={1\over 8}$,
$\beta E_C=1500$ and
$\Delta\tau E_C={1\over 8}$.
The solid lines are predictions \eqref{slsasalk} for the VEVs $\langle\,\re^{\ri q\phi}\rangle_0^2$.
The  dashed  lines include the leading temperature correction\ \eqref{slssaas}
with ${\tilde \mu}_B$ given by \eqref{saslksklkls}.}
\label{fig4}
\end{figure}

We now focus on the considerable 
deviation between 
$\langle\,\re^{\ri q\phi}\rangle_0^2$ and
the MC data for $G_q({\beta\over 2})$
in the domain  $g \gtrsim 0.7$ (see Fig.\,\ref{fig4}).
This discrepancy turns out to be a finite temperature effect  and  can be explored via
infrared (IR) perturbation theory.

To begin,  let us briefly recall some well-known facts 
(see Refs.~\cite{Warner,BLZ2, Lesage} for details)
 concerning the 
low-energy  effective theory for the BSG model \eqref{lksslasas}.
From the RG point of 
view this QFT    describes    a  boundary
flow between two fixed point. Each  fixed point is  controlled by 
the CFT of a free massless Bose field  on the half-line $x\leq 0$ subjected to  
a conformally invariant   boundary
condition at $x=0$.
Whereas the UV behavior   corresponds  to    the free (von Neumann)   boundary condition,
the IR fixed point is governed   by the CFT
with the fixed (Dirichlet)  boundary condition.
To be more precise,
let ${\cal H}^{(n)}$ be the space of states 
of the free   massless Bose field
on the half lane constrained by
the fixed boundary condition 
$\varphi_B={2\pi\over \sqrt{g}}\  n\ (n=0,\,\pm 1\ldots)$.
Then  the whole  space of states of the model \eqref{lksslasas} can be represented as a direct sum
\bea\label{klasosaks}
{\cal H}=\oplus_{n}{\cal H}^{(n)}\ .
\eea
All  terms in this sum  are  isomorphic to each other and
the Hamiltonian associated with   the IR fixed point, 
\bea\label{ksslka}
{\bf H}_*={1\over 4\pi}\ \int_{-\infty}^0\rd x\ \big(\, 
{\boldsymbol\pi}^2+(\partial_x{\boldsymbol\varphi})^2\, \big)\,  \ \ \ \ 
{\rm where}\ 
\ \ \ \ \big[\,{\boldsymbol\pi}(x)\, ,\, {\boldsymbol\varphi}(x')\, \big]=
-2\pi\ri\ \delta(x-x')\ , 
\eea
acts invariantly   in ${\cal H}^{(n)}$,
\bea\label{aslkslsa}
{\bf H}_*\ :\ \ {\cal H}^{(n)}\to {\cal H}^{(n)}\ .
\eea
The effective Hamiltonian  governing  the IR  behavior 
can be introduced  as a certain perturbation by   irrelevant
boundary operators of the critical  Hamiltonian \eqref{ksslka}. 
To describe it explicitly  we note that
the set of  {\it  primary}  boundary fields \cite{Lewellen} of the  
infrared  CFT 
includes  $(\partial_\tau {\tilde {\boldsymbol\varphi}})_B\equiv
\partial_\tau {\tilde {\boldsymbol\varphi}}(0,\tau)$
and 
\bea\label{lakslks}
{\tilde {\boldsymbol {\cal O}}}_m=\big[\,\re^{ {\ri m\over \sqrt {g}}{\tilde {
{\boldsymbol \varphi}}}}\, \big]_B\, , 
\ \ \ \  \ \ \ \  \ \ {\tilde {\boldsymbol {\cal O}}}_m^{\dagger}=
{\tilde {\boldsymbol {\cal O}}}_{-m}
\  \ \ \ \ \ \ \ (\, m=0,\,\pm1,\, \pm 2\ldots\, )\ ,
\eea
where ${\tilde {\boldsymbol \varphi}}$ is the T-dual\footnote{The T-dual of a
free massless field is defined as usual, through the relations:
$\partial_\tau {\tilde {
{\boldsymbol \varphi}}}=\ri\, \partial_x{\boldsymbol \varphi},
\partial_x {\tilde {
{\boldsymbol \varphi}}}=-\ri\, \partial_\tau {\boldsymbol \varphi}.$} 
of ${\boldsymbol \varphi}$.
The   boundary fields\ \eqref{lakslks} 
intertwine the components in the direct sum\ \eqref{klasosaks},
\bea\label{salisaksa}
{\tilde {\boldsymbol {\cal O}}}_m\ \ :\ \  {\cal H}^{(n)}\to {\cal H}^{(n+m)}\, ,
\eea
and  may be normalized
by means of the condition
\bea\label{slsslksa}
\langle\, 
{\tilde {\boldsymbol {\cal O}}}_m(\tau){\tilde {\boldsymbol {\cal O}}}^{\dagger}_m(0)\,
\rangle_0=|\tau|^{-{2m^2\over g}}\ .
\eea
With notation \eqref{lakslks} the   effective Hamiltonian has the form
\bea\label{lsaasksssa}
{\bf H}_{\rm eff}={\bf H}_*-{\tilde \mu}_B\ \big(\,{\tilde  {\boldsymbol {\cal O}}}_1(0)+
{\tilde  {\boldsymbol {\cal O}}^{\dagger}_{1}}(0)\, \big)+\ldots \ .
\eea
Here dots stand for  descendents of
the identity operator of scaling dimensions
$2,\,4,\,6\ldots$
acting  invariantly in each linear subspace ${\cal H}^{(n)}$.
For our purposes the explicit form 
of the omitted terms is not essential, but we shall need an exact  relation
between the IR coupling ${\tilde \mu}_B$ and the original parameter 
${ \mu}_B$  in\ \eqref{lksslasas} \cite{BLZ2, Lesage}:
\bea\label{saslksklkls}
{\tilde \mu}_B=
{\Gamma(1+{1\over g})\over 2\pi}\  \bigg[{2\pi\mu_B\over\Gamma(1+g)}\bigg]^{-{1\over g}}=
{\Gamma(1+\alpha)\over 2\pi}\  \bigg[{2\pi\epsilon
\over\Gamma(1+\alpha^{-1})}\bigg]^{-\alpha}\ \Lambda^{1-\alpha}\ .
\eea

In the IR limit the exponential  boundary fields 
\eqref{lksasaha} should admit asymptotic  expansions through the 
local   boundary fields
from  the CFT  associated with the IR fixed point, i.e.,
\bea\label{ksksals}
{\boldsymbol {\cal O}}_q\asymp \ \sum_{n=-\infty}^{\infty}
\re^{2\pi \ri q n}\  \Big[\, \mu_B^{d_q\over 1-g}\, D(q)\ 
{\boldsymbol I}+\mu_B^{d_q-1\over 1-g}\ D_1(q)\ 
(\partial_\tau {\tilde {\boldsymbol\varphi}})_B+\ldots\, \Big]\ {\mathbb P}_n\ .
\eea
Here ${\mathbb P}_n$  are
projectors on the subspaces ${\cal H}^{(n)}$,  ${\boldsymbol I}$
is the unit operator, and the dots stand for  the contribution of
the conformal  descendents 
of ${\boldsymbol I}$ and $(\partial_\tau {\tilde {\boldsymbol\varphi}})_B$
of scaling dimensions $2,\, 3,\,4\ldots$.
Unfortunately, the explicit form for the  expansion coefficients in \eqref{ksksals}
is currently not known. The sole exception is the function  $D(q)$
given by\ Eq.\,\eqref{laljajsy}.

Using expansions
\,\eqref{lasajlsa},\,\eqref{lsaasksssa},\,\eqref{ksksals}, it is now
straightforward
to calculate the leading temperature 
correction for $G_q(\tau)$\ \eqref{lsajlsajsa}. 
For $\tau={\beta\over 2}$ it is 
given by
\bea\label{slssaas}
G_q\big({\textstyle{\beta\over 2}}\big)=
\langle\,\re^{\ri q\phi}\rangle_0^2\ \bigg[\, 1-4\, \sin^2(\pi q)\  J(\alpha)\ 
{\tilde \mu}_B^2\ \Big({\beta\over 2\pi}\Big)^{2-2\alpha}
+O\big(\,\beta^{-{\rm min}(2, 4\alpha-4)}\,\big)\, \bigg]\ ,
\eea
where the
function $J(\alpha)$ is defined through     analytic
continuation of the integral
\bea\label{skash}
J(\alpha)=\int_{0}^{\pi}\rd u\int_{-\pi}^0\rd v\ \Big[\,4\,\sin^2\big({\textstyle{
u-v\over 2}}\big)\,\Big]^{-\alpha}
\eea
from the  domain of convergence $\Re e\,\alpha< 1$. 
The integral\ \eqref{skash}
can be  expressed in terms of  the generalized hypergeometric function ${}_3F_2$ at unity:
\bea\label{slssl}
J(\alpha)&=&
2^{1-2\alpha } \ \bigg[\,
{\pi^{3\over 2}\Gamma({1\over 2}-\alpha)\over 2\Gamma(1-\alpha) }+
{{}_3F_2\big(1, 1-\alpha, 1; {\textstyle{ 3\over 2}}, 2-\alpha\,|\, 1\big)
\over 1-\alpha}-\\ &&\ \ \ \ \ \ \ \ \ 
{ {}_3F_2\big(1, 1-\alpha, 1-\alpha;
{\textstyle{ 3\over 2}}-\alpha,
 2-\alpha\, |\, 1\big)
\over (1-\alpha)(1- 2\alpha)  }
\bigg]\ .\nonumber
\eea
It should be emphasized that the correction  presented in  Eq.\,\eqref{slssaas}  
 is  the leading one
for $1<\alpha<2$ only. Indeed,
the second  term in  expansion \eqref{ksksals} should   give rise to a
temperature correction  $\propto \beta^{-2}$,  dominating  in the domain
$\alpha>2$. In connection with this, one may   note that
the function $J(\alpha)$ \eqref{slssl} has
a simple pole at $\alpha=2$. It is expected that
the term  $\propto  \beta^{-2}$ 
cancels  out this singularity and makes the expansion \eqref{slssaas} finite 
at $\alpha=2$. Other important features  of $J(\alpha)$ are  that it vanishes at
$\alpha= 1.3917\ldots\  (g=0.7185\ldots)$ and has a simple pole at $\alpha=1$.
For this reason,  one should expect that   the leading low-temperature
correction \eqref{slssaas} can be applied  literally
in  a  narrow vicinity of $\alpha= 1.4$.
The MC results depicted in Fig.\,\ref{fig4}  
are in agreement with this prediction.

\bigskip

\section{$G_q({\beta\over 2})$ in  the vicinity of the
S-I phase transition
\label{Sectree}}
  
One may argue that the low-temperature correction
$\propto ({\tilde \mu}_B)^{2k}\, \beta^{2k(1-\alpha)}$
generated in the perturbation expansion for
the effective Hamiltonian \eqref{lsaasksssa} 
possesses a pole of order  $k$  at $\alpha= 1$. For that reason
the calculation of $G_q({\beta\over 2})$
for  $\alpha$ close to one  requires
an effective  summation of all these singularities.
Below we shall solve this problem 
using the  RG improvement technique.                                                                                                                                      
                                                                                                                                      
\subsection{Running coupling constant}

We begin with an appropriate choice of  running coupling constant 
that provides   a convenient  global ``coordinate'' along
the   boundary flow in the BSG model
including UV and IR fixed points.

In general, the 
definition of the running coupling constant is ambiguous. 
As an example, the RG flow equation\ \eqref{alskask} suggests
the following   simple
choice for  a local coordinate in the vicinity of the UV fixed point:
\bea\label{oaoiaso}
{ f}_{\rm UV}=\Big({ E_*\over E}\Big)^{1-g}\ .
\eea
Here $E$ is the RG energy 
and
 $E_*\propto \mu_B^{1\over 1-g}$ is the integration constant of the RG 
flow equation
which sets the ``physical scale'' in the BSG model.
At the same time  the RG flow arrives at the IR fixed point along
the irrelevant boundary field of scaling dimension $\alpha=g^{-1}$.
Hence
\bea\label{osaksa}
{ f}_{\rm IR}=\Big({ E_*\over E}\Big)^{1-\alpha}
\eea
is another  obvious choice    for the  running coupling.
However,  neither
${ f}_{\rm UV}$ nor   ${ f}_{\rm IR}$
can serve as a  global coordinate for the RG flow, since
they 
fail to describe
the vicinity of the IR and UV fixed points, respectively.

Notice that in this work  we are dealing with   the temperature dependence only, 
so that   the RG  energy $E$ should be understood as  the  temperature and
the dimensionless scaling variable
\bea\label{slslssa}
\kappa={E_*\over E}=\beta E_*
\eea
is the inverse temperature measured in units of the ``physical'' scale.
Also  $E_*/ \mu_B^{1\over 1-g}$
is a dimensionless  RG invariant which
may be an arbitrary  function of
the RG invariant parameter $g$.
It will be convenient for us to  define $E_*$ unambiguously through the relation 
\bea\label{sksaklsaha}
E_*=
{1\over 2\pi}\ \bigg[{2\pi { \mu}_B
\over \Gamma(1+g)}\bigg]^{1\over 1-g}=
{\Lambda\over 2\pi}\ \bigg[{2\pi \epsilon\over \Gamma(1+g)}\bigg]^{1\over 1-g}\ .
\eea

\begin{figure}[t]
\centering
\includegraphics[width=10cm]{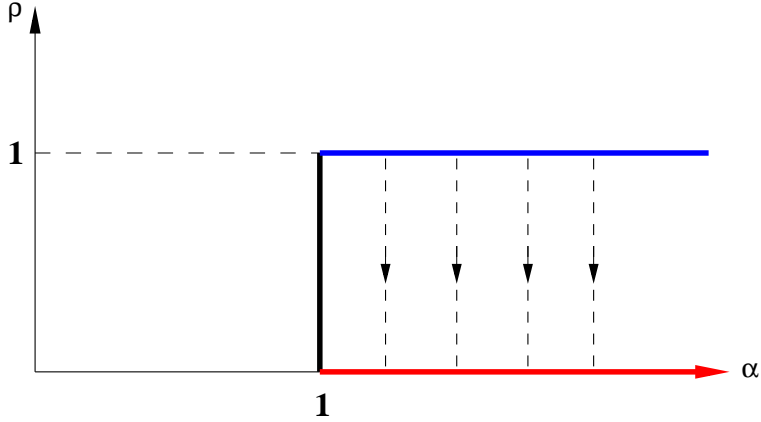}
\caption{The RG  flow for  the BSG model
in the $(\alpha,\rho)$ plane. 
The orientation of  the RG trajectories (dashed lines)
corresponds to a decrease of the RG energy $E$ (temperature). 
The bold rays $\rho=0,\, 1$ and the bold  segment at
$\alpha=1$ are  lines of fixed points.}
\label{fig5}
\end{figure}

It turns out that 
the scaling function ${\rho}={\rho}(\kappa)$ \eqref{lakssoisiul}
is  a  convenient  choice for the running coupling
for the boundary flow in the  BSG model.
As was mentioned in the introduction,  this function is known exactly.
It was originally
calculated  
for
integer values of  $\alpha=2,\,3,\, 4\ldots$ within the Thermodynamic
Bethe Ansatz approach in  Ref.\,\cite{Saleur}. Later  the closed expression for
${\rho}$ was found for  arbitrary  $\alpha>1$\ \cite{BLZ}.
That   result   can be summarized as follows.
Let $\Psi$ be the solution
of the ordinary differential equation
\bea\label{lslalsa}
\Big[-\partial^2_y+s\ \re^{gy}+\re^y\ \Big]\ \Psi=0
\eea
such that
\bea\label{lsjlss}
\Psi(y)\to \begin{cases}
0\, ,\ \ &y\to+\infty\\
const\ ( y-y_0)\,,\ \ &y\to-\infty\ .
\end{cases}
\eea
Then  ${ \rho}$ can be  expressed in terms of 
the function $y_0=y_0(s)$\ \eqref{lsjlss}:
\bea\label{slsjlsa}
{\rho}(\kappa)=1+g\ s\,{\partial y_0\over\partial s}\ ,
\eea
provided
\bea\label{jsgtaoiu}
s=g^2\ \kappa^{2-2 g}\ .
\eea
With the exact result\ \eqref{slsjlsa} one can  
prove (see \cite{BLZ} for details)
that ${ \rho}$ admits the following high- and low-temperature  expansions
\bea\label{lsaskssa}
{ \rho}(\kappa)=
1-\sum_{n=1}^{\infty}\rho_n(\alpha)\ ({ f}_{\rm UV})^{2n}\,\asymp\,
\sum_{n=1}^{\infty}{\tilde  \rho}_n(\alpha)\ ({ f}_{\rm IR})^{2n}\ .
\eea
Here the expansion in terms of the UV running coupling constant
${ f}_{\rm UV}=\kappa^{1-g}$ 
has a finite radius of convergence, 
while the power series in ${ f}_{\rm IR}=\kappa^{1-\alpha}$ is an asymptotic
expansion with zero radius of convergence. 
The coefficients in these  series are 
functions of $\alpha$   satisfying  a remarkable duality relation
originally conjectured by Schmid \cite{Schmid}:
\bea\label{lkslsaj}
\rho_n(\alpha)={\tilde  \rho}_n(\alpha^{-1})\ .
\eea
The first coefficient ${\tilde  \rho}_1$ explicitly reads\ \cite{BLZ2}
\bea\label{lsasals}
{\tilde  \rho}_1(\alpha)=2^{1-2\alpha}\ \sqrt{\pi}\ \, {\Gamma^3(1+\alpha)\over
\Gamma({1\over 2}+\alpha)}\ .
\eea
As follows from Eqs.\,\eqref{lsaskssa}-\eqref{lsasals},  
both ${ f}_{\rm UV}^2$ and ${ f}_{\rm IR}^2$
can be perturbatively expressed in terms of $1-\rho$ and $\rho$, 
respectively, 
for any value of $\alpha$. 
Hence the pair  $(\alpha,\,\rho)$ may serve as 
global coordinates for  the  RG  flow in the BSG model   (see Fig.\,\ref{fig5}).

It is instructive to examine the RG flow equation in terms of the 
coordinates $(\alpha,\,\rho)$. 
Using Eqs.\,\eqref{lsaskssa} and\, \eqref{slsjlsa}
it is possible to show that
\bea\label{lsasaasiu}
E\ {{\rm d} \alpha\over  {\rm d} E}&=&0\ ,\nonumber \\
E\ {{\rm d} { \rho}\over  {\rm d} E}&=&{2(\alpha-1)\,
{  \rho} (1-{ \rho})\over 1+(\alpha-1){ \rho}}\ 
\bigg[\, 1-{(\alpha-1)^2\over\alpha}\ U(\alpha,{ \rho})\, \bigg]\ .
\eea 
Here  
$U({\alpha, \rho})$ is a  bounded function\footnote{Our numerical results 
suggest that $0< U(\alpha,{ \rho})< {\rm min}\big(\,1,
{\textstyle  {\alpha\over (\alpha-1)^2}}\, \big)$.} 
 for $0\leq\alpha,\,\rho\leq 1$ which
admits the following
convergent and asymptotic expansions
\bea\label{sklsasakj}
U({ \rho},\alpha)=
\sum_{m=1}^{\infty}U_m(\alpha)\ (1-{  \rho})^m\, ,\ \ \ \ 
U({  \rho},\alpha)
\asymp \sum_{m=1}^{\infty}{\tilde U}_m(\alpha)\, {  \rho}^m\ .
\eea
Notice that  the expansion  
coefficients satisfy the duality relation $U_m(\alpha)={\tilde U}_m(\alpha^{-1})$.
It is an immediate consequence of \eqref{lsasaasiu} that
\bea\label{salsjls}
{\rho}={\bar  \rho}+O\big(\, (\alpha-1)^2\,\big)\ ,
\eea
where
${\bar { \rho}}$ is the solution of the differential equation,
\bea\label{slkslsa}
E\, {{\rm d} {\bar { \rho}}\over  {\rm d} E}=
{2(\alpha-1)\, 
{\bar { \rho}} (1-
{\bar { \rho}})\over 1+(\alpha-1){\bar {\rho}}}\ ,
\eea
satisfying the asymptotic condition 
${\bar { \rho}}
\to\ {\tilde \rho}_1\  (\beta E_*)^{2-2\alpha}$ as $E=\beta^{-1}\to 0$ with
${\tilde \rho}_1$ given by  \eqref{lsasals}.
One can  easily integrate the differential equation \eqref{slkslsa} and
show that  ${\bar { \rho}}$ solves the  equation
\bea\label{laslaksla}
{\bar { \rho}}\ (1-{\bar { \rho}})^{-\alpha}=
2^{1-2\alpha}\ \sqrt{\pi}\ {\Gamma^3(1+\alpha)\over
\Gamma({1\over 2}+\alpha)}\ (\beta E_*)^{2-2\alpha}\ \ .
\eea

Despite the fact
that the function  ${\bar { \rho}}$ 
defined   by Eq.\eqref{laslaksla}
does not have a simple physical meaning,  like 
the ``physical'' coupling   ${ \rho}$\ \eqref{lakssoisiul},
it may play the role  of  a running coupling   in the same sense 
as  ${ \rho}$.
In fact in the domain of our interest, i.e. $1\leq \alpha \lesssim 1.4$,
one can neglect the difference between 
${ \rho}$ and ${\bar { \rho}}$ 
since the resulting  numerical error
turns out   to be
smaller than the error bars of the MC data.

\subsection{MC results for ${R\over R_S}$
\label{subsecFourB}}

Numerically,
we compute the zero-bias resistance by extrapolating the Fourier
transform $\langle\, \phi \phi\, \rangle_{\omega_k}$ of $\langle\, \phi(\tau)\phi(0)\, \rangle$
in  Matsubara frequencies  $\omega_k={2\pi k\over\beta}\to 0$:
\bea
\frac{R}{R_S}={\alpha\over2\pi}\ \Big[\,
|\omega_k|\,\langle\, \phi \phi\,\rangle_{\omega_k}\,
\Big]_{\omega_k\to 0}
\approx {\alpha\over N^2}\ \Big[\,
|k|\ \langle\, |{\tilde \phi}_k |^2\, \rangle\, \Big]_{\omega_k\to 0}\ \ ,
\label{zero_bias_res}
\eea
where we used the notation \eqref{fouriercoefficient}.
It is therefore sufficient to sample the
modules $|\tilde \phi_k |^2$ of the lowest few Fourier coefficients.
\begin{figure}[ht]
\centering
\includegraphics[width=10cm, angle=-90]{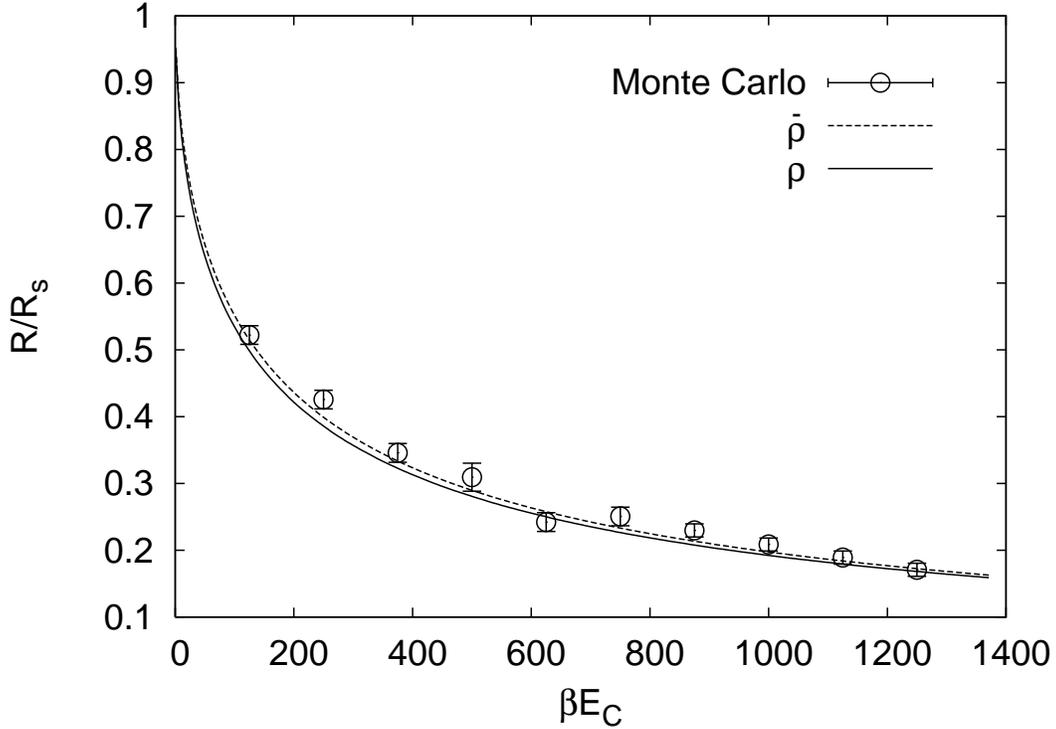}
\caption{  ${R\over R_S}$
as a function of $\beta E_C$.
The MC data were obtained   for $\alpha=1.4$,  ${E_J\over E_C}={1\over 4}$ and
$\Delta\tau E_C={1\over 8}$. The solid and dashed lines represent
${ \rho}$\ \eqref{slsjlsa} and
${\bar { \rho}}$\ \eqref{laslaksla}, respectively with
$E_*$ given by 
Eqs.\,\eqref{sksaklsaha},\,\eqref{kjsshkjsa},\,\eqref{lslsaas}.}
\label{fig6}
\end{figure}

\begin{figure}[ht]
\centering
\includegraphics[width=10cm, angle=-90]{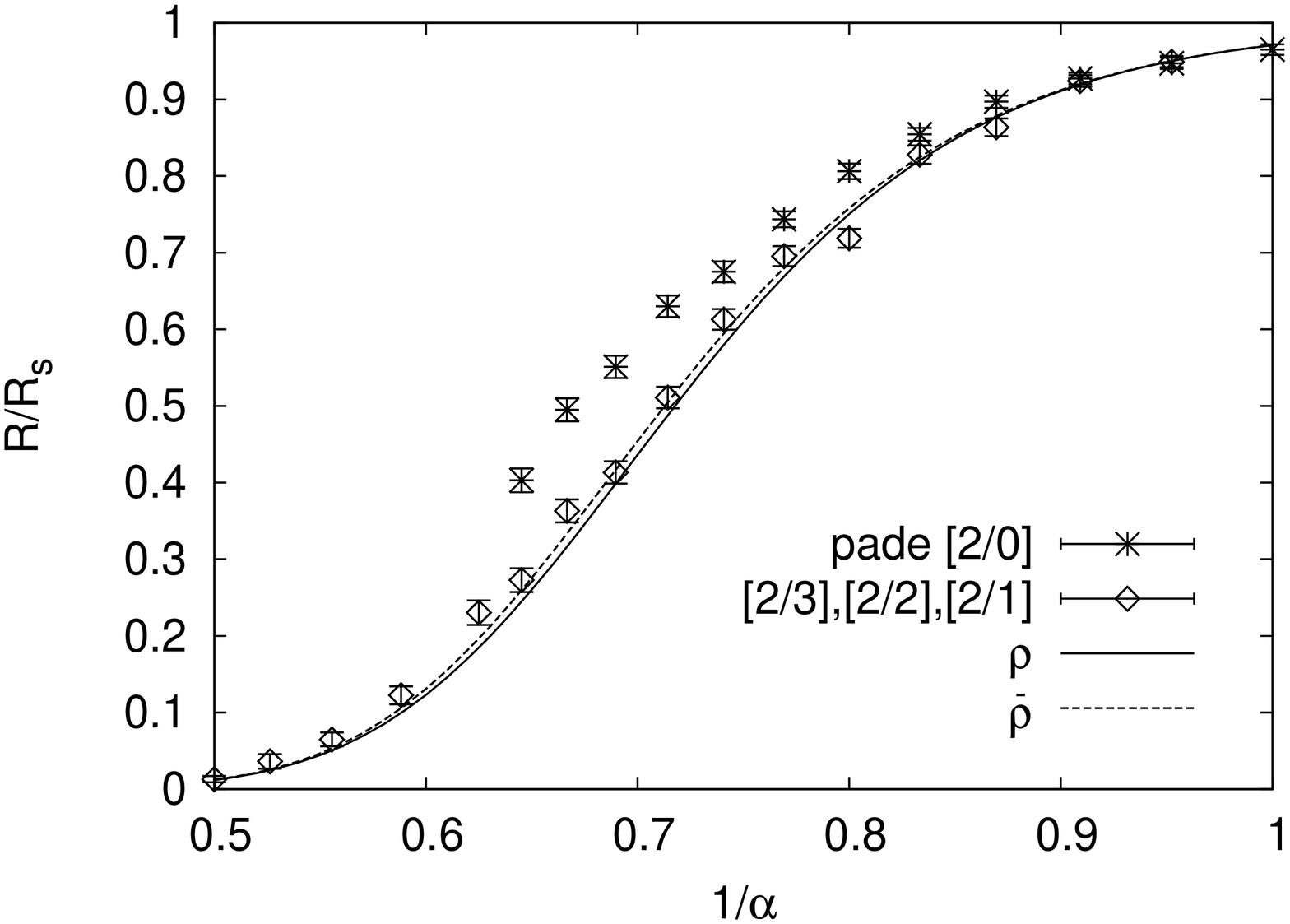}
\caption{  ${R\over R_S}$
as a function of $\alpha$.
The MC data were obtained  
 for $\beta E_C=1500$,  ${E_J\over E_C}={1\over 8}$ and
$\Delta\tau E_C={1\over 8}$
(the same  as in
Fig.\,\ref{fig4}). The solid and dashed lines represent
${ \rho}$\ \eqref{slsjlsa} and
${\bar { \rho}}$\ \eqref{laslaksla}, respectively, with
$E_*$ given by 
Eqs.\,\eqref{sksaklsaha},\,\eqref{kjsshkjsa},\,\eqref{lslsaas}.}
\label{fig7}
\end{figure}

In Fig.\,\ref{fig6}  we
compare the MC results for the ratio ${R\over R_S}$
as a function of $\beta E_C$  for $\alpha=1.4$ and $E_J/E_C={1\over 4}$   
against
${ \rho}$\ \eqref{slsjlsa} and ${\bar { \rho}}$
\eqref{laslaksla} with $E_*$ given by 
Eqs.\,\eqref{sksaklsaha},\,\eqref{kjsshkjsa}.
Figure~\ref{fig7} shows the analytical prediction 
and numerical results for ${R\over R_S}$
as a function of $\alpha$.
Here, the MC data were obtained   
for $\beta E_C=1500$,  ${E_J\over E_C}={1\over 8}$ and
$\Delta\tau E_C={1\over 8}$, i.e., for
the same values of parameters as in 
Fig.\,\ref{fig4}.

Although the cluster algorithm allows us 
to study temperatures much lower 
than what was previously accessible 
(which means that the frequency points are densely spaced), 
different extrapolation procedures 
can yield quite different results for $R/R_S$. 
A parabolic fit ([2/0] Pad\'e) to the lowest 
9 Matsubara points gives the result 
shown by stars in Fig.\,\ref{fig7}. 
The downturn at the lowest frequencies is 
better captured by higher 
order Pad\'e functions. 
We tried [2/1], [2/2] and [2/3] Pad\'e approximations 
to the lowest 9 frequency
points and found similar results for $R/R_S$. 
These estimates (shown as diamonds in Fig.~\ref{fig7}) 
agree reasonably well with the 
theoretical prediction for $\rho$ and $\bar\rho$.
 
\bigskip

\subsection{Anomalous dimensions \label{subsecFourC}}

Once we have chosen the 
global coordinates 
for the boundary RG flow,
the scaling functions should be regarded as functions of $(\alpha,\rho)$.
Here, our prime interest is the
scaling limit  of  the two-point correlator,
\bea\label{lissas}
\lim_{\epsilon\to 0\atop\beta\Lambda\to\infty }\epsilon^{-{2gq^2\over 1-g}}\,
G_q\big({\textstyle{\beta\over 2}}\big)= 
{\cal G}_{q}(\alpha,\rho)\  \ \ \ \ \ {\rm for}\ \ \ \ 
 |q|\leq {\textstyle {1\over 2}}\ .
\eea

Let us first consider the case $|q|< {\textstyle {1\over 2}}$.
The renormalization theory
predicts the general structure of 
the   perturbative  expansions of ${\cal G}_{q}$. Namely,  
for $|q|< {\textstyle {1\over 2}}$ the exponential boundary operator
is renormalized multiplicatively and, therefore,
\bea\label{kjssksk}
{\cal G}^{(\rm pert)}_{q}(\alpha,\rho)= 
Z_q^2(\alpha,{ \rho})\, W_q(\alpha,{ \rho})\ ,\eea
where 
the anomalous dimension,
\bea\label{slsjslsa}
\gamma_q(\alpha,{ \rho})={E\over Z_q}\ {\rd Z_q\over \rd E}\ ,
\eea
and the factor
$W_q$   possess   (formal) 
power series  expansions in $\rho$ and $1-\rho$:
\bea\label{slslalalssi}
W_q(\alpha,{ \rho})&=&\sum_{k=0}^\infty  w_k\
(1- \rho)^k=
\sum_{k=0}^\infty {\tilde w}_k\ 
{ \rho}^k\ ,\nonumber\\
\gamma_q(\alpha,
{ \rho})&=&\sum_{k=0}^\infty  \gamma_k\
(1- \rho)^k =\sum_{k=0}^\infty 
{\tilde \gamma}_k\ { \rho}^k\ .
\eea
It is important to bear in mind that
both $W_q$ and $\gamma_q$ are defined ambiguously, since
the transformation $W_q\to  W_q\, F^2$, $Z_q\to Z_q/F$, i.e.,
\bea\label{slsajslaa}
\gamma_q\to \gamma_q-
\beta_{\rho} \ {\partial \log F\over \partial \rho}\ \ \ \ \ \ \ \ 
\Big(\, \beta_\rho =E\, {{\rm d} { \rho}\over {\rm d} E}\, \Big)\, ,
\eea
preserves
${\cal G}^{(\rm pert)}_{q}$. 
Here  $F$ is  an arbitrary function in a rather broad class
possessing a  formal power series expansions
at $\rho=1$ and $\rho=0$.
This  ambiguity  does not affect
the anomalous dimension at the critical lines ${ \rho}=0,\,1$ and
$\alpha=1$ (see Fig.\,\ref{fig5}). 
At the line of  UV fixed points
${ \rho}=1$,
the operator 
${\boldsymbol {\cal O}}_q$ becomes  the   primary conformal boundary
field of   scaling 
dimension $d_q$\ \eqref{klaksa}.
On the other hand (as follows from Eq.\,\eqref{ksksals})
${\boldsymbol {\cal O}}_q$  becomes  the unit 
operator in each subspace ${\cal H}^{(n)}$
in the IR limit. Thus, we conclude that
\bea\label{slsajsa}
\gamma_q(\alpha,1)={q^2\over\alpha}\, ,\ \ \ \ \ \ \ \ \ 
\gamma_q(\alpha,0)=0\ .
\eea
Thanks to Refs.~\cite{Polchinski,Thorlacius} 
the value of the anomalous dimension at  the S-I phase transition
line
$\alpha=1$ is known (see also
Section\,\ref{Sectfour} for some explanations):
\bea\label{lassalsao}
\gamma_q(1,  \rho)={1\over\pi^2}\ \arcsin^2\big(\sqrt{ \rho}\,
\sin(\pi q)\, \big)\ .
\eea
The inherent ambiguity \eqref{slsajslaa} in the definition of
the anomalous dimension
outside  the critical region
allows one to choose   $\gamma_q$ of the form
\bea\label{lsoassuia}
\gamma_q(\alpha,  \rho)={1\over\alpha \pi^2}\ \arcsin^2\big(\sqrt{ \rho}\,
\sin(\pi q)\, \big)\ ,
\eea
which admits local
expansions
\eqref{slslalalssi} and consistently fulfills the global restrictions
\eqref{slsajsa}, \eqref{lassalsao}.

All the above considerations suggest to resum the perturbative 
$\big($\,in 
$f^2_{\rm IR}\propto{\tilde \mu}^2_B\beta^{2-2\alpha}\,\big)$
part  of  expansion \eqref{slssaas}   and, using 
Eqs.\,\eqref{lsaskssa},\,\eqref{lsasals},
express it
in the form 
\bea\label{osiuslssaas}
G^{\rm (IR)}_q\big({\textstyle{\beta\over 2}}\big)&=&
\langle\,\re^{\ri q\phi}\rangle_0^2\ \
\exp\bigg[\, {2\over\alpha \pi^2}\ \int_{\beta E_*}^\infty{\rd \kappa\over 
\kappa}\ \arcsin^2\big(\sqrt{ \rho(\kappa)}\,
\sin(\pi q)\, \big)
\bigg]\times \nonumber 
\\ && \bigg[\,1+\sum_{k=1}^\infty {\tilde w_k}\, \rho^k\, \bigg]\ ,
\eea
where
\bea\label{ajslsa}
{\tilde w_1}&=&
- {\sin^2(\pi q)\over \pi^2}\ 
\bigg\{{2^{2\alpha-1}\Gamma({1\over 2}+\alpha)
\over \sqrt{\pi}\Gamma(1+\alpha)}\ J(\alpha)+
{1\over \alpha (\alpha-1)}\, \bigg\}=\nonumber 
\\&& - {\sin^2(\pi q)\over \pi^2}\ \log 4+
O\big(\,(\alpha-1)^1\,\big)\ .
\eea
The major advantage of  
\eqref{osiuslssaas} is that
all  singularities at   $\alpha=1$ are now  absorbed
by the exponential factor, i.e.,
the expansion coefficients  ${\tilde w}_k$
are regular functions of $\alpha$
in the vicinity of the S-I phase transition  line.

Similarly, to
study  $G_q({\textstyle{\beta\over 2}})$
in the vicinity  $\rho=1$, it makes sense  to
write  its high-temperature  perturbative expansion in the form,
\bea\label{rqelsslssa}
G^{\rm (UV)}_q\big({\textstyle{\beta\over 2}}\big)&=&
\bigg[\,1+\sum_{k=1}^{\infty}c_k(q)\,\epsilon^{2k}\,\bigg]\
\exp\bigg[\, {(-2)\over\alpha \pi^2}\ \int^{\beta E_*}_{\pi
E_*\over \Lambda}{\rd \kappa\over
\kappa}\ \arcsin^2\big(\sqrt{ \rho(\kappa)}\,
\sin(\pi q)\, \big)
\bigg]\times \nonumber
\\ && \bigg[\,1+\sum_{k=1}^\infty w_k\ (1-\rho)^k\, \bigg]\ .
\eea
Here the overall multiplicative  constant
should be chosen to satisfy the condition
\bea\label{lsalsalsa}
G^{\rm (UV)}_q\big({\textstyle{\beta\over 2}}\big)\to
\Big({\beta\Lambda\over \pi}\Big)^{-2d_q}\  \big[\,C^{(0)}(q,\epsilon)\, \big]^2\
\ \ \ \ \ \ \ {\rm as}\ \ \ \ \ \
\beta\Lambda\gg 1\gg \beta E_*\to 0\ ,
\eea
which follows from the
expansion \eqref{lasajlsa}. In particular,
the first  coefficient  $c_1(q)$ reads 
\bea\label{ssalsakasl}
c_1(q)=2C_1^{(0)}(q)+{4\pi q\tan(\pi q)\over \alpha-1}\ {\sqrt{\pi}
\Gamma(1+\alpha^{-1})\over 2\Gamma({1\over 2}+\alpha^{-1})}\ ,
\eea
where $C_1^{(0)}(q)$ is calculated in the appendix (see Eq.\,\eqref{askjsa}).
Recall that
the integration measure
${{\rm d}\kappa\over \kappa}$ in Eqs.\,\eqref{osiuslssaas},\,\eqref{rqelsslssa}
is expressed through the running constant
by means of the RG flow equations \eqref{lsasaasiu}, i.e.,
\bea\label{alasslkassa}
{\rd\kappa\over\kappa}=-{1+(\alpha-1)\rho\over 2(\alpha-1)\rho (1-\rho)}\ \rd\rho
+
O\big(\,(\alpha-1)^1\,\big)\ .
\eea

The values  $q=\pm {1\over 2}$ require  special consideration.
In this case it is useful
to start with  the bare  boundary fields
\bea\label{okslsssa}
{ {\cal O}}_+=\cos\big({\textstyle{\phi\over 2}}\big)\ ,\ \ \ \ \ \ 
{ {\cal O}}_-=\sin\big({\textstyle{\phi\over 2}}\big)\ .
\eea
They  carry  the same charge with respect to the   
abelian group ${\mathbb Z}$   \eqref{lskls}.
Nevertheless the ${ {\cal O}}_\pm$  do not mix under    RG transformations
since
they have different ``C-parities''
with respect
to the  flip    $\phi\to -\phi$.
Thus, the bare  boundary fields ${ {\cal O}}_\pm$ are
renormalized multiplicatively 
with the corresponding anomalous
dimensions  $\gamma_\pm(\alpha,\rho)$. Whereas 
these anomalous 
dimensions   are the same at the critical line $\rho=1$,
\bea\label{amsssii}
\gamma_\pm(\alpha,1)={1\over 4\alpha}\ ,
\eea
they are different in the  IR limit. 
Due to  C-parity conservation
one can predict that   ${ {\cal O}}_+$ and ${ {\cal O}}_-$
renormalize   in the IR limit   to  the unit
operator ${\boldsymbol I}$
and $\partial_\tau{\tilde {\boldsymbol \varphi}}$ 
(in each subspace ${\cal H}^{(n)}$ \eqref{klasosaks}), respectively,
and hence
\bea\label{osiusamsssii}
\gamma_+(\alpha,0)=0\ , \ \ \ \ \ \ \gamma_-(\alpha,0)=1\ .
\eea
An explicit form  of $\gamma_\pm(\alpha,\rho)$ at the critical line $\alpha=1$
follows again from  results of  Refs.\,\cite{Polchinski,Thorlacius} 
(see Section\,\ref{Sectfour}):
\bea\label{lsajslasa}
\gamma_+(1,  \rho)={1\over \pi^2}\ \arcsin^2(\sqrt{ \rho}\,
)\ ,\ \ \ \ \ \ \ \ 
\gamma_-(1,  \rho)=\bigg[\, {\arcsin(\sqrt{ \rho}\,
)\over\pi}-1\,\bigg]^2\ .
\eea
Using Eqs.\,\eqref{amsssii},\,\eqref{osiusamsssii} and \eqref{lsajslasa}
one can  propose  the following   expressions for $\gamma_\pm$: 
\bea\label{pjslsaas}
\gamma_+(\alpha,  \rho)&=&\gamma_{1\over 2}(\alpha,  \rho)=
{1\over \alpha\pi^2}\ \arcsin^2(\sqrt{ \rho})\ ,\\
\gamma_-(\alpha,  \rho)&=&1-{2\over\pi}\ \arcsin(\sqrt{ \rho})+
{1\over \alpha\pi^2}\ \arcsin^2(\sqrt{ \rho})\ .\nonumber
\eea

The correlator $G_{1\over 2}({\beta\over 2})$ is expressed 
through the two-point  correlation functions of 
${ {\cal O}}_\pm$\ \eqref{okslsssa}:
\bea\label{lakslsasa}
G_{1\over 2}\big({\textstyle{\beta\over 2}}\big)=
G_+\big({\textstyle{\beta\over 2}}\big)+
G_-\big({\textstyle{\beta\over 2}}\big)
\, ,\ \ \ \ \ \ {\rm where}\ \  \ \ G_{\pm}(\tau)=
\langle\, {\cal O}_\pm(\tau){\cal O}_\pm(0)\, \rangle\ .
\eea
Notice that the anomalous   dimensions $\gamma_+$ and  $\gamma_-$
are well separated in the IR limit (see Eq.\,\eqref{osiusamsssii}) 
and the term 
$G_-\big({\textstyle{\beta\over 2}}\big) \propto (\beta E_*)^{-2}$
in \eqref{lakslsasa}
is negligible in the low-temperature limit.
For this reason,  Eq.\,\eqref{osiuslssaas}
can be  applied for $q=\pm {1\over 2}$ without any modification.
At the same time the high-temperature perturbative 
expansion\ \eqref{rqelsslssa} is not applicable
for $|q|$ sufficiently close to ${1\over 2}$. For $q=\pm {1\over 2}$
it
should be replaced by \eqref{lakslsasa} where $G_{\pm }$ are
understood as
\bea\label{rqelssa}
G^{\rm (UV)}_{\pm }
\big({\textstyle{\beta\over 2}}\big)&=&{ {1\over 2}}\
\bigg[\,1+\sum_{k=1}^{\infty}c^{(\pm)}_{k\over 2}\epsilon^k\,\bigg]\
\exp\bigg[ -2\int^{\beta E_*}_{\pi
E_*\over \Lambda}{\rd \kappa\over
\kappa}\, \gamma_\pm \big(\alpha, \rho(\kappa)\big)
\bigg]\times\nonumber \\ &&
\bigg[\,1+\sum_{k=1}^\infty w^{(\pm)}_{k\over 2}\
(1-\rho)^{k\over 2}\, \bigg]\ .
\eea
It is
straightforward to calculate the first  nontrivial
expansion coefficients
in  \eqref{rqelssa}:
\bea\label{lsksjslsaas}
w^{(\pm)}_{1\over 2}&=&\pm {\Gamma({\alpha-1\over 2\alpha})\over
\sqrt{\pi}\Gamma(1-{1\over 2\alpha})}\
\bigg[\,{2^{2/\alpha}\, \Gamma({1\over 2}+\alpha^{-1})\over
2\sqrt{\pi}\Gamma(1+\alpha^{-1})}\,\bigg]^{1\over 2}\mp {2\alpha\over \pi
(\alpha-1)}+
{2\over\pi}=\nonumber\\
&&\pm{\log 4-1\over \pi}+{2\over\pi}+O\big((\alpha-1)^1\big)\ ,
\eea
and
\bea\label{slsslqoqui}
c^{(\pm)}_{1\over 2}=\pm 2\, C^{(1)}_0\big({\textstyle{1\over 2}}\big)-
4\ \bigg[\, 1\mp {\alpha\over \alpha-1}\, \bigg]
\ \bigg[{\sqrt{\pi}\Gamma(1+\alpha^{-1})\over
2\Gamma({1\over 2}+\alpha^{-1})}\bigg]^{1\over 2}\ ,
\eea
with $C^{(1)}_0(q)$  given by \eqref{aslasksala}.

\begin{figure}[t]
\centering
\includegraphics[width=10cm, angle=-90]{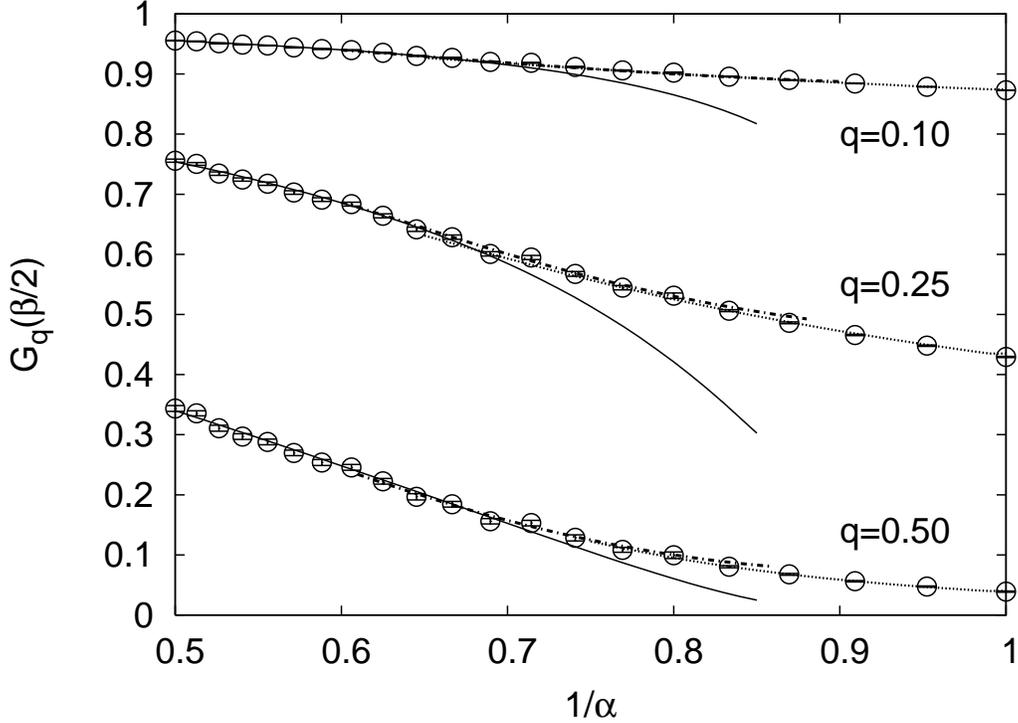}
\caption{
$G_q({\beta\over 2})$ as a function of $\alpha^{-1}$
for  $q=0.1,\, 0.25,\, 0.5$.
The MC data were obtained  for ${E_J\over E_C}={1\over 8}$,
$\beta E_C=1500$ and
$\Delta\tau E_C={1\over 8}$.
The solid lines are predictions 
\eqref{slsasalk} for the VEVs $\langle\,\re^{\ri q\phi}\rangle_0^2$.
The dashed-dotted lines correspond
to the RG-improved low-temperature expansion\ 
\eqref{osiuslssaas},\,\eqref{ajslsa}.
The dotted lines represent 
the  RG-improved high-temperature expansions
\eqref{rqelsslssa} for  $q=0.1,\,0.25$ and
\eqref{lakslsasa}-\eqref{slsslqoqui} for $q=0.5$.
All the terms in  the  perturbative series in $1-\rho$ in \eqref{rqelsslssa}, 
except the first one, were neglected.}
\label{fig8}
\end{figure}

In Fig.\,\ref{fig8}
the   MC data   from Fig.\,\ref{fig4}
are compared against
the RG-improved low-temperature expansion\
\eqref{osiuslssaas},\,\eqref{ajslsa} and the 
RG-improved high-temperature expansions
\eqref{rqelsslssa},\, \eqref{ssalsakasl} for  $q=0.1,\,0.25$ and
\eqref{lakslsasa}-\eqref{slsslqoqui} for $q=0.5$. 
Notice that all
the terms in  the  perturbative series in $1-\rho$ in \eqref{rqelsslssa}
were neglected except the first one.
For  the  calculation of  the integrals  in
\eqref{osiuslssaas},\,\eqref{rqelsslssa},\,\eqref{rqelssa}
we employed   Eq.\,\eqref{alasslkassa}.
Similar results were found for $\beta E_C=375$ and $750$.

\section{Scaling dimensions of $\re^{\ri q\phi}$ at $\alpha=1$ \label{Sectfour} }

As follows from the RG flow equation\ \eqref{alskask}, 
to  achieve the scaling limit in the case $g=\alpha^{-1}=1$
the bare coupling
constant $\epsilon$ \eqref{lslsaas} should be kept fixed.
Hence, the limiting value $\rho$ \eqref{ajsksajas} is a certain function
of the dimensionless ratio ${E_J\over E_C}$. It   is
discussed in the next section.
Here we just  mention that $\rho$ is a monotonically decreasing function
of ${E_J\over E_C}$\ (see Fig.\,\ref{fig11}) and, hence, the variable $0<\rho<1$ 
can be used to  parametrize the critical line $\alpha=1$
to  the same extent  as  ${E_J\over E_C}$.
Notice that 
the commonly used coupling for  the BSG model at $g=1$ is the one introduced
in Ref.\,\cite{Maldacena}.
To relate  $\rho$ to   the 
Calan-Klebanov-Ludwig-Maldacena
coupling, $g_{CKLM}$, one  needs to repeat
the calculation from  the work \cite{Lin} 
in the renormalization scheme of \cite{Maldacena}. This yields
\bea\label{sjaisiasu}
\sqrt{\rho}=\cos\big(\pi g_{CKLM}\big)\ .
\eea

The one-parameter 
family of  boundary CFT associated with the line of fixed points $\alpha=1$
possesses the global  ${\mathbb Z}$-invariance\ \eqref{lskls}. For this reason
the corresponding 
boundary fields  can be labeled by
the ``quasi-charge''
\bea\label{ksjskjs}
-{\textstyle{1\over 2}} < {\hat q} \leq{\textstyle{1\over 2}}
\eea
(analogous to the quasi-momentum of Bloch states in a periodic
lattice), and
the spectrum of their   scaling dimensions
splits into bands.
Let us denote  the    primary boundary
fields by  ${\boldsymbol {\cal O}}_{({\hat q}, m)}$,
where
${\hat q}$ is  the quasi-charge from the first Brillouin zone
\eqref{ksjskjs} and the integer $m=0,\,\pm1,\,\pm 2\ldots$  labels
the different bands.
According to the works \cite{Polchinski,Thorlacius}  the scaling dimension of
${\boldsymbol {\cal O}}_{({\hat q}, m)}$
is given by
\bea\label{slsjlsahst}
d_{({\hat q},m)}=\big(\lambda({\hat q})+m\big)^2\ ,
\eea
where
\bea\label{sasaksa}
\lambda({\hat q})={\textstyle {1\over\pi}}\
\arcsin\big(\sqrt{\rho}\, \sin(\pi {\hat q})\big)
\eea
lies within the range
\bea\label{slssloui}
-{\textstyle  {1\over \pi}}\ 
\arcsin(\sqrt{\rho})< 
\lambda\leq {\textstyle  {1\over \pi}}\ \arcsin(\sqrt{\rho})\ .
\eea
Thus, the allowed values of $\lambda({\hat q})+m$
consist of bands of width ${2\over\pi}\,\arcsin(\sqrt{\rho})$
centered at  every integer, with gaps of width $1-{2\over\pi}\,\arcsin(\sqrt{\rho})$.

Similar to Eq.\,\eqref{lasajlsa},
one should expect that 
the bare exponential operators for $\alpha=1$ admit expansions of the form: 
\bea\label{lsalsasa}
\re^{\ri q\phi}(\tau)\asymp \sum_{m= 0,\pm 1,\pm 2\ldots} C^{(m)}({\hat q},\rho)\ 
\Lambda^{-d_{({\hat q},m)}}\ 
{\boldsymbol {\cal O}}_{({\hat q}, m)}(\tau)+\ldots\, ,\ \,  
{\rm with}\ \,  {\hat q}+{\textstyle{1\over 2}}=
q+{\textstyle{1\over 2}}\, ({\rm mod}\,{\mathbb Z})
\eea
and the dots stand  for the conformal descendents of the scaling
dimension $d_{({\hat q},m)}+N$\ $(N=1,\,2,\ldots)$.
The immediate consequence of \eqref{lsalsasa} is the 
expansion for  the two-point correlator\ \eqref{lsajlsajsa}:
\bea\label{lsslsassa}
G_q(\tau)\asymp \sum_{m= 0,\pm 1,\pm 2\ldots\atop
N=0,\,1,\,2\ldots}\ A^{(m)}_N\ \ 
\Big({\beta\Lambda\over \pi}\Big)^{-2d_{({\hat q},m)}-2N}\ \  
\Big(\,\sin\Big|{\pi \tau\over\beta}\Big|\, \Big)^{-2d_{({\hat q},m)}-2N}\ .
\eea
Here the dimensionless  coefficients  $A^{(m)}_N$ are some
functions of  ${\hat q}$ and $\rho$.
As long as $|q|<{1\over 2}$ and $\beta\Lambda\gg 1$,  the first  term with $m=N=0$ dominates
in this  asymptotic   series:
\bea\label{slssalsaiiu}
G_q(\tau)\,\approx\, A^{(0)}_0 \ \ \Big({\beta\Lambda\over \pi}\Big)^{-2\eta_q}\ \
\Big(\,\sin\Big|{\pi \tau\over\beta}\Big|\, \Big)^{-2\eta_q}\ \ \ \ \ \ \ \ \ 
\big(|q|<{\textstyle {1\over 2}}\big)\ ,
\eea
where $\eta_q=d_{(q ,0)}$ \eqref{ssalkiuy}. Notice  that, as follows
from Eqs.\eqref{slsjlsahst},\,\eqref{sasaksa},
$d_{({1\over 2} ,0)}=d_{({1\over 2} ,-1)}$ for $\rho=1$.
Hence, near the boundary ${\hat q}={1\over 2}$
of the first Brillouin zone \eqref{ksjskjs} 
and for $\rho$ close to one, it
makes sense to 
keep also the first subleading term with $m=-1,\, N=0$ in  \eqref{lsslsassa}.
In particular,
\bea\label{salsaiiu}
G_{1\over 2}(\tau)\,\approx\, A^{(0)}_0 \ \ 
\Big({\beta\Lambda\over \pi}\Big)^{-2\eta}\ \
\Big(\,\sin\Big|{\pi \tau\over\beta}\Big|\, \Big)^{-2\eta}+
A^{(-1)}_0 \ \
\Big({\beta\Lambda\over \pi}\Big)^{-2\eta'}\ \
\Big(\,\sin\Big|{\pi \tau\over\beta}\Big|\, \Big)^{-2\eta'}\ ,
\eea
where 
\bea\label{laksskl}
\eta=\eta_{ {1\over 2}}\, ,\ \ \ \ \ \ \ \ \ \ \ \ 
\eta'=\big(\,1-\sqrt{\eta_{ {1\over 2}}}\,\big)^2\ .
\eea
One may note that 
the limit
$\alpha\to 1$ brings 
 the UV expansions \eqref{rqelsslssa}  and 
\eqref{lakslsasa},\, \eqref{rqelssa}
into the forms  \eqref{slssalsaiiu} and \eqref{salsaiiu},
respectively, taken at $\tau=\beta/2$. 

\begin{figure}[t]
\centering
\includegraphics[width=10cm, angle=-90]{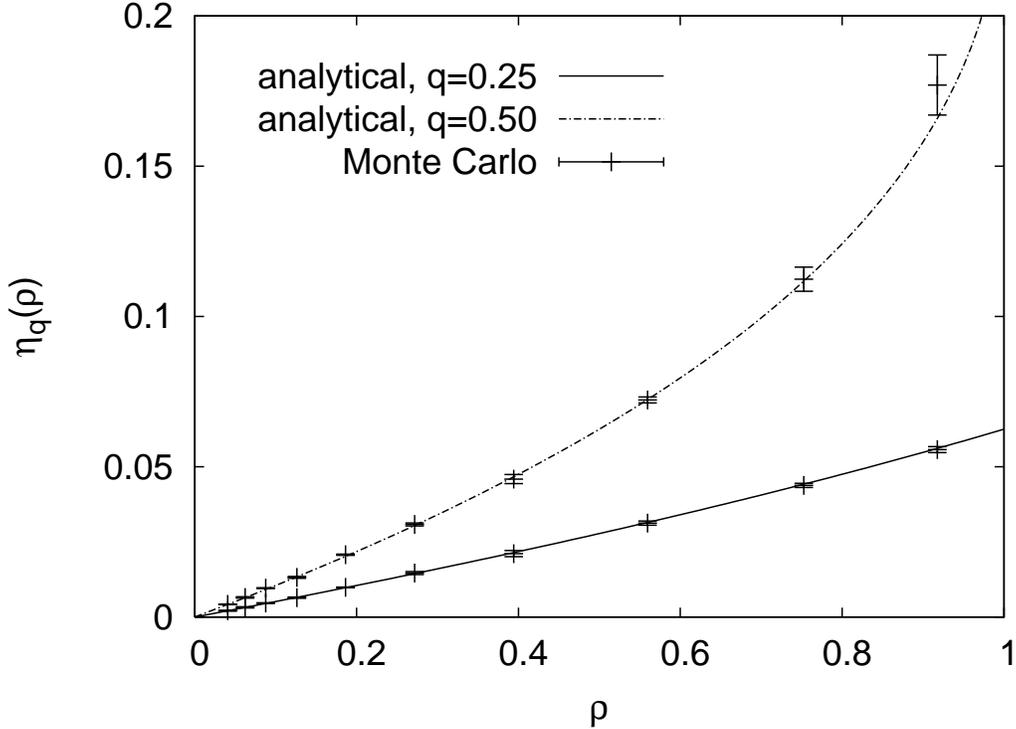}
\caption{The critical exponent
$\eta_q$ as a function of $\rho$ \eqref{ajsksajas}.
The MC data were obtained  for 
$\beta E_C=375$,
$\Delta\tau E_C={1\over 8}$, ${E_J\over E_C}=0.6, 1.0, 1.2, \ldots 2.0$, and $\beta E_C=750$,
$\Delta\tau E_C={1\over 8}$, ${E_J\over E_C}=0.2,\,0.4, 0.8$ (first, second and fourth point from the right). Errors on $\rho$ are smaller than the symbol size. 
The lines show the predictions of Eq.\,\eqref{ssalkiuy}.}
\label{fig9}
\end{figure}
\bigskip

We have performed the numerical check of Eqs.\,\eqref{slssalsaiiu} and 
\eqref{salsaiiu},\,\eqref{laksskl}.
Namely, we used these general 
forms to fit the function $G_q(\tau)$ obtained from 
the MC simulations. The constants $A^{(0)}_0, A^{(-1)}_0$ and $\eta_q$ 
were treated as  fitting parameters. 
In Fig.\,\ref{fig9} the results of these fits for  $\eta_q$
are compared against 
the analytical prediction\ \eqref{ssalkiuy} for $q=0.25,\, 0.5$
and several values of $\rho$.

\begin{figure}[t]
\centering
\includegraphics[width=10cm, angle=-90]{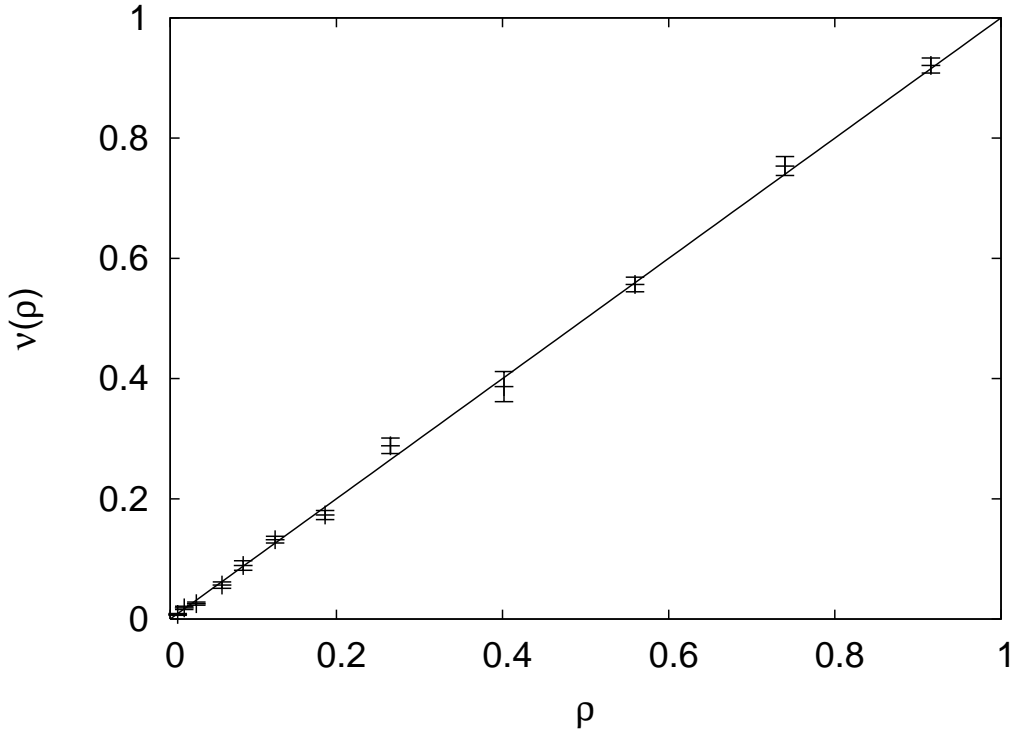}
\caption{Coefficient 
$\nu$\ \eqref{lslsj} as a function of $\rho$\ \eqref{ajsksajas}.
The points   were obtained by fitting the MC data   for
$\beta E_C=375,\, 750,\, 1000,\, 1250,\, 1500$ with
$\Delta\tau E_C={1\over 8}$ and (from right to left) ${E_J\over E_C}=
0.2, 0.4, 0.6, 0.8, 1, 1.2, 1.4, 1.6, 1.8, 2.2, 2.6, 3$. 
Errors on $\rho$ are smaller than the symbol size. 
The solid line is $\nu=\rho$.}
\label{fig10}
\end{figure}

We have also  numerically   studied the mean phase fluctuation for $\alpha=1$. 
As it  was observed
in Ref.\,\cite{WT}, the phase fluctuations grow proportional
to the logarithm of the inverse temperature,
\bea\label{lslsj}
\langle\,(\phi-{\bar\phi})^2\, \rangle\to 2\nu\ \log(\beta E_C)+const\ 
\ \ \ \ \ \ \ \ \ \ \ \ (\beta E_C\to\infty)\ .
\eea
In Fig.\,\ref{fig10} our  numerical  results for $\nu$, 
obtained by  fitting the form\ \eqref{lslsj} to the MC data for 
$\langle\,(\phi-{\bar\phi})^2\,\rangle$,  are plotted against $\rho$.
In all likelihood,
\bea\label{lsajslsa}
\nu=\rho={{1\over 2}}\ 
{ \partial^2 d_{q,0}\over \partial q^2}\Big|_{q=0}\ .
\eea

\bigskip

\section{\label{Sectfive}Nonperturbative energy scale} 

Scaling functions like $\rho$ \eqref{lakssoisiul} and
${\cal G}_q$ \eqref{lissas}  are defined
as a particular  limit of  their corresponding ``bare'' counterparts from
the Caldeira-Leggett model.
However, as  has been demonstrated above,
they can be successfully
employed  for numerical approximations
of  the  bare
functions for nonzero  $\epsilon$ and finite values of   
$\beta\Lambda$ provided 
the energy scale $E_*$ is understood as in\ \eqref{sksaklsaha}.
Of course,
the bare functions differ from their scaling limits
by  corrections   vanishing as $\epsilon\to 0$
and the   quality of  the   ``scaling''  approximation 
diminishes as  the bare coupling  $\epsilon$ is increased.
It turns out that  some of these
corrections can be captured, and
the overall  quality of  the scaling approximation considerably improved,  
by substituting $E_*$ with some nonperturbatively defined
``physical'' scale   for  the Caldeira-Leggett model.
Recall that   $E_*$  has been    introduced 
as a  first integral of the
RG flow  equation\ \eqref{alskask}, i.e., it is a
perturbatively defined   RG-invariant energy scale in
the  BSG model.

To introduce the physical scale   in the 
Caldeira-Leggett model, $E_*^{\rm (CL)}$, we take advantage of
the result of Ref.\,\cite{WG}. Namely, Weiss and Grabert
found that at low temperatures
\bea\label{lasjaiq}
{R\over R_S}\asymp {\pi^2\sqrt{\pi}\, \Gamma(1+\alpha)\over
2\, \Gamma({1\over 2}+\alpha)}\ \Big({\beta\omega_c\over \pi}\Big)^{2-2\alpha}\ 
\Big({\Delta\over \omega_c}\Big)^2+\ldots \ ,
\eea
where 
\bea\label{alksk}
\omega_c=\sqrt{8E_J E_C}
\eea
and $\Delta$ is the so-called 
tunneling amplitude (see also book \cite{Weiss}).
It is important that  the validity of the leading
low-temperature asymptotic \,\eqref{lasjaiq} does not actually
require the smallness of the ratio ${E_J\over E_C}$, and
therefore, the nonperturbative energy scale  can be
identified as 
\bea\label{lkjlssa}
E_*^{\rm (CL)}={\omega_c\over 2\pi}\ \  
\bigg[\,{\pi \Delta \over \omega_c\,
\Gamma(1+\alpha)}\, \bigg]^{1\over 1-\alpha}\ .
\eea
It is easy to see that 
Eq.\,\eqref{lasjaiq}, when  expressed
in terms of  $E_*^{\rm (CL)}$,
has  the  form
of  the leading low-temperature correction to the scaling function $\rho$
\eqref{lsaskssa},\,\eqref{lsasals}
provided $E_*$ is replaced by $E_*^{\rm (CL)}$. 
One should expect that 
the scaling function $\rho$\ \eqref{slsjlsa}
with $\kappa$ replaced by $\beta E_*^{\rm (CL)}$ well approximates
${R\over R_S}$ even for 
${E_J\over E_C}\gtrsim 1$, provided the temperature is not too
high compared to the cutoff scale $\Lambda\propto\alpha E_C$.
By virtue of Eqs.\,\eqref{kjsshkjsa},\,\eqref{lslsaas} and
\eqref{sksaklsaha},
one has
\bea\label{alksls}
E_*^{\rm (CL)}=
{E_C\over 2\Gamma(1+\alpha^{-1})}\ \  
\bigg[
{\pi^2\, \re^{-\gamma_E}
\over 4\Gamma(\alpha^{-1})}\ \bigg]^{1\over
\alpha-1}\ \ \ \big[\, z\, F(z,\alpha)\, \big]^{\alpha\over
\alpha-1}\, , 
\eea 
where $F$ is some function defined for any 
\bea\label{lsalsklsa}
z\equiv {E_J\over E_C}\geq 0\ ,
\eea
such that $F(0,\alpha)=1$.

The function $F=F(z,\alpha)$ in Eq.\eqref{alksls}  can be 
studied using
conventional 
perturbation theory in the bare coupling. In particular, at the
second order in perturbation theory 
we find 
\bea\label{lsklsas}
F( z, \alpha)=1+ f_1(\alpha)\ z^2+
O\big(\,z^{{\rm min}(\,4, {\alpha+2 \over \alpha-1}\, )}\, \big)\ ,
\eea
where
\bea\label{kasjsksa}
f_1(\alpha)={\textstyle {1\over 2}}\ C_1^{(0)}(1)\ \ \Big({\pi 
\re^{-\gamma_E}\over 8\alpha}\Big)^2\ ,
\eea
with $C_1^{(0)}(1)$ the coefficient $C_1^{(0)}(q)$\ \eqref{askjsa}
taken at $q=1$.
Notice that  $f_1(\alpha)$ has a simple pole at $\alpha=4$
which is expected to be canceled    by the nonperturbative
correction $\propto z^{\alpha+2 \over \alpha-1}$, and, therefore, 
the term  presented in Eq.\,\eqref{lsklsas} 
is a leading small-$z$ correction in the domain
$1\leq \alpha<4$.
Numerical values for  $f_1(\alpha)$ for
$1/1.2\leq \alpha\leq 2$ are presented in Table\,\ref{kiiajsh}.
\begin{table}[ht]
\begin{center}
\begin{tabular}{| r | l | l |}
\hline
\rule{0mm}{4mm}
$g$\hspace{3mm}  &\ $\alpha$\hspace{1mm}
  &\ \ \  $f_1$ \\
\hline
\hline
\rule{0mm}{3.6mm}
0.50  & 2.00 &   0.11513  \\
0.60  & 1.67 &   0.13243 \\
0.70  & 1.43 &   0.15176 \\
0.80  & 1.25 &   0.17177\\
0.90  & 1.11 &   0.19203\\
1.00  & 1.00 &   0.21234\\
1.10  & 0.91 &   0.23260\\
1.20  & 0.83 &   0.25278\\
\hline
\hline
\end{tabular}
\end{center}
\caption{ Coefficient  $f_1$\ \eqref{kasjsksa} 
as a function of $\alpha=g^{-1}$.}
\label{kiiajsh}
\end{table}

In the  strong coupling limit, ${E_J\over E_C}\to \infty$, the
energy scale $E_*^{\rm (CL)}$ can be explored by means of the
saddle-point
approximation and
the original  instanton calculation from Ref.~\cite{Schmid} suggests   the following semiclassical form
for the tunneling amplitude $\Delta$ in\ \eqref{lkjlssa}:
\bea\label{lksasklsasa}
{\Delta\over 2 \omega_c}={\textstyle\sqrt{2\over \pi}}\ 
\ (8z)^{1\over 4}\  \ \re^{-\sqrt{8z}}\ 
\Big[\, 1+O\big(z^{-{1\over 2}}\big)\,  \Big]\ .
\eea

\begin{figure}[ht]
\centering
\includegraphics[width=10cm, angle=-90]{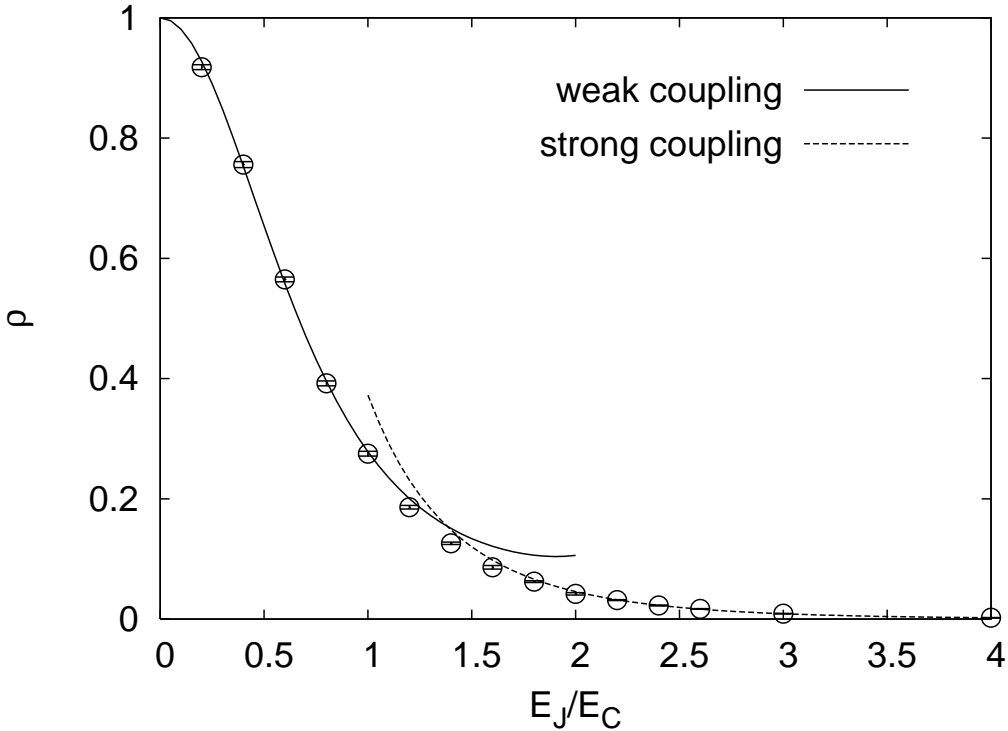}
\caption{
$\rho$\ \eqref{ajsksajas}
as a function of ${E_J\over E_C}$ for $\alpha=1$.
The MC data were obtained for $\beta E_C=375$ and
$\Delta\tau E_C={1\over 8},\, {1\over 16}$.
The solid and dashed lines    correspond 
to   weak  coupling  \eqref{alssasksu} and
strong coupling  \eqref{alksu}  expansions, respectively, with
the numerical coefficients   chosen as
$f_1=0.21234\ldots$, $f_2\approx-0.05$ and 
$a\approx 0.5$.}
\label{fig11}
\end{figure}

To illustrate the results from this section,    let us  consider
the $E_J/E_C$-dependence of $\rho$ 
at $\alpha=1$. Due to the relation \eqref{salsjls}, 
one can apply  Eq.\eqref{laslaksla} 
with $E_*$ replaced by  $E_*^{\rm (CL)}$\ \eqref{lkjlssa},\,\eqref{alksls}
 as $\alpha\to 1$.
Then, for $\alpha=1$ one finds:
\bea\label{ksasksa}
\rho={1\over 1+X^2}\, \ \ \ \ \ \  \ {\rm where}\ \ \ \ \ 
X={\omega_c\over \pi\Delta}={\pi^2\over 4\re^{\gamma_E}}\ z\ F(z, 1)\ .
\eea
We may now apply  
the weak and strong coupling expansions 
\eqref{lsklsas}, \eqref{lksasklsasa} to 
the case $\alpha=1$:
\bea\label{alssasksu}
\rho^{(\rm weak)}={1\over 1+X^2}\ ,\ \ \ 
X={\pi^2\over 4\re^{\gamma_E}}\  \ z\
 \big[\, 1+f_1\, z^2+
f_2\, z^4+\ldots\, \big]\, \ \ \ \ \ \ 
\big(\, z={\textstyle {E_J\over E_C}}\ll 1\, \big)\, ,
\eea
and
\bea\label{alksu}
\rho^{(\rm strong)}=8\pi \ (8z)^{{1\over 2}}\
\re^{-2\sqrt{8z}}\ \Big[\,1+a\, z^{-{1\over 2}}+O(z^{-1})\, \Big]
\, \ \ \ \ \ \ \big(\, z={\textstyle {E_J\over E_C}}\gg 1\, \big)\, .
\eea
The numerical value of the coefficient $f_1$ in \eqref{alssasksu}
is presented in
Table\,\ref{kiiajsh}:
\bea\label{alsasal}
f_1\big|_{\alpha=1}=0.21234\ldots\ .
\eea
Coefficients $f_2$ \eqref{alssasksu} and $a$\ \eqref{alksu}
are currently  not known, and we have
estimated their values using  the available   MC data
(see Table\,\ref{kiiajshsksju} and Fig.\,\ref{fig11}):
\bea\label{slsals}
f_2\approx -0.05\ ,\ \ \ \ \ \ \ a\approx 0.5\ .
\eea

\begin{table}[ht]
\begin{center}
\begin{tabular}{| r | l |l |l | }
\hline
\rule{0mm}{4mm}
${E_J\over E_C}$\hspace{1mm} &\ \ \ \ \ \ \ \ $\rho$
 & $\rho^{\rm (weak)} $ & $\rho^{\rm (strong)} $ \\
\hline
\hline
\rule{0mm}{3.6mm}
0.2  & 0.918 $\pm$ 0.004 &    0.9276   &      \\
0.4  & 0.753 $\pm$ 0.005 &    0.753  &      \\
0.6  & 0.565 $\pm$ 0.004 &    0.558  &      \\
0.8  & 0.392 $\pm$ 0.004 &    0.396    &      \\
1.0  & 0.275 $\pm$ 0.004 &    0.278   &     \\
1.2  & 0.186 $\pm$ 0.003 &    0.20 &    \\
1.4  & 0.126 $\pm$ 0.002 &    0.15    & 0.15    \\
1.6  & 0.086 $\pm$ 0.003 &    0.12    & 0.098   \\
1.8  & 0.0622 $\pm$ 0.0015 &          & 0.066   \\
2.0  & 0.0420 $\pm$ 0.002  &          & 0.046    \\
2.2  & 0.0315 $\pm$ 0.0008 &          & 0.032    \\
2.4  & 0.0226 $\pm$ 0.0008 &          & 0.0228    \\
2.6  & 0.0168 $\pm$ 0.0005 &          & 0.0164  \\
3.0  & 0.0090 $\pm$ 0.0004 &          & 0.0088  \\
4.0  & 0.0022$\pm$ 0.0002 &           & 0.0022 \\
\hline
\hline
\end{tabular}
\end{center}
\caption{ $\rho$ \eqref{ajsksajas} 
as a function of ${E_J\over E_C}$ for $\alpha=1$.
The MC data were obtained for $\beta E_C=375$ and
$\Delta\tau E_C={1\over 8},\, {1\over 16}$.
The expansions
$\rho^{\rm (weak)} $ and $\rho^{\rm (strong)} $ 
are given by \eqref{alssasksu} and
\eqref{alksu}, respectively, with
the numerical coefficients chosen as
$f_1=0.21234\ldots$, $f_2\approx-0.05$ and
$a\approx 0.5$. 
  }
\label{kiiajshsksju}
\end{table}

\bigskip

\section{\label{six}Conclusion}

The resistively shunted Josephson junction,
represented by the Caldeira-Leggett action with non-compact
phase variable, is one of the simplest and most fundamental
models  exhibiting a dissipation driven quantum phase transition.
It has been studied theoretically for more than twenty years. Most recently,
there has been considerable interest in phenomena attributed to the
interplay between spatial couplings and dissipation in systems
involving several junctions \cite{Refael03, WRT} and Josephson
junction arrays \cite{Tewari06, Goswami06, Refael07}.
Complicated phase diagrams and lines of fixed points
have been predicted for the latter models using
strong- and weak-coupling approximations.

The purpose of the present paper was to highlight 
some  analytical results for the
single junction model, which -- at least 
in the solid state community -- have not
been widely appreciated. 
We took advantage of the theoretical machinery developed in
the  QFT/String Theory  community to discuss
 analytical expressions characterizing the behavior of the junction
at criticality and in the superconducting phase.
In particular, we studied the Matsubara  correlator
$\langle\, \re^{\ri q\phi}(\tau)\, \re^{-\ri q\phi}(0)\,\rangle$ as a
function of $\alpha$, $q$,  $E_J/E_C$ and temperature. On the S-I phase boundary,
this correlator decays with a critical exponent,
which is a
function of the dimensionless  zero-bias resistance $\rho$.
We also discussed  an implicit expression for $\rho$ as a function
of $\alpha$ and  temperature.
The second quantity of interest to us was
the mean phase fluctuation $\langle (\phi-\bar\phi)^2\rangle$. At the critical
line $\alpha=1$, the fluctuations grow
proportional to the logarithm of the inverse
temperature, 
$\langle (\phi-\bar\phi)^2\rangle\rightarrow 2\nu \log(\beta E_C)+const$
as $\beta\rightarrow \infty$, and we have  proposed that $\nu=\rho$.

The accuracy of the analytical formulas and the
range of applicability of the perturbative calculations
have been tested by means of MC simulations.
These numerical checks were made possible by the recent development
of an efficient cluster sampling scheme \cite{WT}. Hence, the remarkable
agreement between analytical and numerical results not
only verifies the accuracy of the formulas discussed in this paper,
but also demonstrates the performance of the cluster algorithm.

\section*{Acknowledgments}

The MC simulations were performed on 
the Hreidar beowulf cluster at ETH Z\"urich, using the 
ALPS library \cite{ALPS}. 
We thank M. Troyer for the generous allocation of computer time.
SL also acknowledges     useful discussions with
L.B. Ioffe,  Al.B. Zamolodchikov and
A.B. Zamolodchikov.

\bigskip

\noindent Research work of SL  is supported in part by DOE grant
$\#$DE-FG02-96 ER 40949. PW acknowledges support from NSF-DMR-040135.

\appendix
 
\section{Appendix}
In the case of the Caldeira-Leggett model, 
the coefficients in the perturbative series  \eqref{slsaskjsa}
do not possess a simple analytic form.
Here, we  consider  integral representations for $C_1^{(0)}(q)$ and $C_0^{(\pm 1)}(q)$
which are sufficient for their  numerical evaluations.

The perturbative  propagator for the action \eqref{taction} with $E_J=0$ explicitly reads
\bea\label{lsssls}
\langle\, \phi(\tau)\phi(0)\, \rangle_{\rm pert}={4\pi g\over\beta}\ \sum_{n=1}^{\infty}
{\cos(\omega_n\tau )\over \omega_n+{\omega^2_n\over\Lambda}\, \re^{\gamma_E}}\ .
\eea
Here $ \omega_n={2\pi n\over
\beta}$ are    Matsubara frequencies and  notations \eqref{lkss},
\eqref{kjsshkjsa} are applied.
As the temperature goes to zero, one has
\bea\label{ksasak}
\langle\, \phi(\tau)\phi(0)\, \rangle_{\rm pert}=2g\, \log\big(
{\textstyle{\Lambda\beta\over 2\pi}}\big)+
\log \mathfrak{G}(\Lambda\tau)+
O\big({\textstyle{1\over \Lambda\beta}}\big) \ ,
\eea
with
\bea\label{slkslsaa}
 \mathfrak{G}(v)=\exp\bigg[\,
-4g  \int_{0}^{\infty}\rd\nu\ {\sin^2({\nu v\over 2})\over
\nu +\nu^2\ \re^{\gamma_E}}\ \bigg]\ .
\eea
Notice that
the zero-temperature Green function \eqref{slkslsaa}  satisfies  the 
conditions $\mathfrak{G}(0)=1$ and 
\bea\label{laaslsa}
 \mathfrak{G}(v)= |v|^{-2g}\, \big(1+O(v^{-2})\, \big)\ \ \ \ \ \ {\rm as}\ \ \ |v|\to\infty\ .
\eea

For $\Re e (qg)>{1\over 2}$ the perturbative coefficient $C_0^{(1)}(q)$ is given by
\bea\label{lalsassq}
C_0^{(1)}(q)= \int_{-\infty}^{\infty}\rd v\ \mathfrak{G}^q(v)\ .
\eea
The coefficient $C_1^{(0)}(q)$ also admits a simple  integral representation
\bea\label{akasj}
C_1^{(0)}(q)=\int_{-\infty}^{\infty}\rd v\int_{-\infty}^{\infty}\rd u
\big[\, \mathfrak{ G}(v-u)\, \mathfrak{ G}^q(v)\mathfrak{ G}^{-q}(u)-
\mathfrak{ G}(v-u)\, \big]
\eea
within the domain of convergence
\bea\label{ksksksa}
g>1\ ,\ \ \ \ \ -q_0< \Re e\, (q)<q_0\, \ \ \ \ \ {\rm with}\ \ \
 q_0=1-{\textstyle {1\over 2g}}\ .
\eea

Both  integrals\ \eqref{lalsassq},\,\eqref{akasj}  
diverge in the  domain of our interest, i.e. for
$0<g<1$ and $|q|\leq {1\over 2}$.
Such divergences are of a similar nature as those
encountered in calculations of  the structure functions
within conformal perturbation theory\ \cite{ALZ}.
Namely, they  indicate the presence  of  nonperturbative
terms in the $\epsilon$-expansion \eqref{lasa}, i.e., 
this is a price we have to pay for the analyticity of
the renormalization constants\ \eqref{slsaskjsa}.
To assign an appropriate meaning to expressions
\eqref{lalsassq} and \eqref{akasj} in
the domain of  interest one should regularize the integrals somehow and
then extract their  finite parts.
Of course, the  finite parts are not fixed unambiguously  
 by any general principle.
In  a sense,  this implies  an  ambiguity   in the  choice
of the bare coupling.
As is usual in QFT,
in the absence of logarithmic divergences (which is the case)
the  analytical regularization (an analytic continuation in  field
dimension)   proves to 
be  the most convenient for  calculations.
In the case under consideration it 
implies a certain  renormalization scheme based
on the simplest admissible  RG flow equation\ \eqref{alskask}.

For the above reasons 
we introduce  $C_0^{(1)}(q)$ and $C_1^{(0)}(q)$
through an  analytic continuation in  $g$ (the scaling dimension of 
the perturbation in \eqref{lksslasas})
of the integrals\ \eqref{lalsassq},\, \eqref{ksksksa}
from their domains of convergence.
In the case of  $C_0^{(1)}(q)$ the analytic continuation
can be easily performed by 
rewriting \eqref{lalsassq} in the form
\bea\label{aslasksala}
C_0^{(1)}(q)=\int_{-\infty}^{\infty}\rd v\ \big[\,
\mathfrak{G}^q(v)-\mathfrak{\bar G}^q(v)\, \big]+{\bar C}_0^{(1)}(q)\ ,
\eea
where
\bea\label{aaksala}
{\bar C}_0^{(1)}(q)=
{\sqrt\pi \Gamma(gq-{1\over 2})\over \Gamma(g q)}
\eea
and
\bea\label{klaasksl}
 \mathfrak{\bar G}(v)=
(1+v^2)^{-g}\ .
\eea
Now the integral in \eqref{aslasksala} converges for $\Re e (gq)>-{1\over 2}$
and
defines an analytic  function of the variable $g q$.
This implies in particular that
 $C_0^{(1)}(1+q)$, considered
as a  function of the complex variable $q$  for  $0<g<1$
has a simple pole at $q=-q_0={1\over 2g}-1$  
with the residue:
\bea\label{saalaaiu}
C_0^{(1)}(1+q)\to {g^{-1}\over q+q_0}\ \ \ \ \ {\rm as}\ \ \ q\to-q_0\ .
\eea
Notice that the coefficient $C_0^{(-1)}(q)$ can be calculated
using \eqref{aslasksala} and the relation $C_0^{(-1)}(q)=C_0^{(1)}(-q)$.

An analytic 
 continuation of  the  integral \eqref{akasj}  is slightly more subtle.
Before proceeding with its explicit description  let us  make an obvious   remark concerning 
the function \eqref{klaasksl}. 
It is easy to see that\ $\log \mathfrak{\bar G}(\Lambda\tau)$\ can be thought of as  
a perturbative, zero-temperature 
propagator for the model defined by the Matsubara action
\bea\label{lsaslas}
{\bar {\mathscr   A}}={1\over 4\pi g\beta}\ 
\sum_{n=-\infty}^{\infty} |\omega_n|\ \re^{{|\omega_n|\over\Lambda}}\ |{\bar \phi}(n)|^2-E_J\
 \int_{-{\beta\over 2}}^{\beta\over 2}\rd\tau \,
\cos(\phi)\ ,
\eea
where ${\bar \phi}(n)= \int_{-{\beta\over 2}}^{\beta\over 2}\rd\tau\, \phi(\tau)\ \re^{\ri \omega_n\tau}$.
Despite the 
lack of  a  physical interpretation, the action  is   a suitable  UV
regularization of the boundary effective action for the  QFT\ \eqref{lksslasas}.
Furthermore it has 
pleasant advantages compared to the Caldeira-Leggett action\ \eqref{taction}.
Namely, the  first perturbative coefficients
in the 
expansions of renormalization constants for \eqref{lsaslas}
(which will be denoted as ${\bar C}^{(n)}_k(q)$) admit
relatively simple 
analytic forms. The coefficient
${\bar C}_0^{(1)}(q)$ is given by \eqref{aslasksala}, while
${\bar C}_1^{(0)}(q)$ can be expressed in terms
of the generalized hypergeometric function ${}_4F_3$:
\bea\label{kskhhshs}
 {\bar C}_1^{(0)}(q)&=&
 {2^{3 - (1-q) g}\, \pi^{5\over 2}\, \Gamma(-{1\over 2}+(1-q) g)
\over \cos(\pi q g) (1-g) \Gamma(qg)}\
\times \nonumber \\
&&
 {}_4F_3^{\rm reg}\big(
-{\textstyle {1\over 2}},
-1 + g, -q\, g,
-{\textstyle {1\over 2}}
+(1-q)\, g \,;\,
{\textstyle {1\over 2}}  - q\, g\, ,
-{\textstyle {1\over 2}}+   {\textstyle {1-q\over 2}}\, g\, ,\,
   {\textstyle {1-q\over 2}}\, g\,    \big|\, {\textstyle {1\over 4}} \big) -\nonumber \\ &&
{2^{1 - (1+q) g} \, \pi^{5\over 2}\ \Gamma(-{1\over 2}+ (1+q)g)\over
\cos(\pi q g) \Gamma(-q g)}\times \\ &&
{}_4F_3^{\rm reg}\big(\,
 {\textstyle {1\over 2}}\,,\,
 g\,,\, q\, g\,, -{\textstyle {1\over 2}} + (1+q)\, g\,;\,
     {\textstyle {1+q\over 2}} \, g\, ,\,
{\textstyle {3\over 2}} + q\, g\,,\,
{\textstyle {1\over 2}}+{\textstyle {1+q\over 2}}\, g\, \big|\,
{\textstyle {1\over 4}}\, \big)\ ,
\nonumber
\eea
where
\bea\label{aslkss}
&&{}_4F_3^{\rm reg}(a,b,c,d; e,f,g\,| z)=
{1\over \Gamma(e)\Gamma(f)\Gamma(g)}\times\\ &&
\ \ \ \ \ \ \ \ \ \  \bigg[\, 1+{a b c d\over ef g}\ {z\over 1!}+
{a(a+1) b (b+1) c (c+1) d (d+1)\over e (e+1)f (f+1) g(g+1)}\ {z^2\over 2!}+\ldots\, \bigg]\ .\nonumber
\eea

To write down an  efficient integral representation for 
$  C_1^{(0)}(q)$ in the Caldeira-Leggett model we first observe 
that $ C_1^{(0)}(q)$ has three
simple poles located at the boundary of
the domain of  convergence\ \eqref{ksksksa}:
\bea\label{laslsjlsa}
C_1^{(0)}(q)\to-{2\pi q\tan(\pi g q)\over 1-g}\ \ \ {\rm as }\ \ \
g\to 1\ ,
\eea
and
\bea\label{lsajls}
C_1^{(0)}(q)\to  C_0^{(1)}(1\mp q) C_0^{(1)}(\pm q)
\ \ \ {\rm as }\ \ \
\pm q\to q_0=1-{\textstyle{1\over 2g}}\ ,
\eea
where $C_0^{(1)}(q)$  is defined  by Eq.\,\eqref{aslasksala} and  satisfies  \eqref{saalaaiu}.
This observation implies that for numerical evaluation of $ { C}_1^{(0)}(q)=
{ C}_1^{(0)}(-q)$
in the domain $0\leq q\leq 1$ and $g>0$ it makes sense 
to rewrite \eqref{akasj}
in the form,
\bea\label{askjsa}
C_1^{(0)}(q)={\bar C}_1^{(0)}(q)+\delta C_1^{(0)}(q)\ ,
\eea
where
\bea\label{lsalsa}
&&\delta C_1^{(0)}(q)={ C}_0^{(1)}(-q) { C}_0^{(1)}(1+q)-{\bar C}_0^{(1)}(-q) {\bar  C}_0^{(1)}(1+q)+
\int_{-\infty}^{\infty}\rd v \Big\{\\ &&
{ C}_0^{(1)}(1-q) \mathfrak{ G}^{q}(v)-{ C}_0^{(1)}(-q) \mathfrak{ G}^{1+q}(v)+
{\bar C}_0^{(1)}(-q) \mathfrak{\bar  G}^{1+q}(v)-{ \bar C}_0^{(1)}(1-q)\mathfrak{\bar G}^{q}(v)+\nonumber\\ && 
\int_{-\infty}^{\infty}\rd u\ 
\big[\, \mathfrak{ G}(v-u)\, \mathfrak{ G}^q(v)\mathfrak{ G}^{-q}(u)-
\mathfrak{ G}(v-u)-\mathfrak{ G}^{q}(v) \mathfrak{ G}^{1-q}(u)-
\big(\, \mathfrak{ G}\to \mathfrak{\bar G}\, \big)\, \big]\,\Big\}\ .\nonumber 
\eea
It should be noted that for  the numerical
results depicted in Figs.\,\ref{fig2} 
and \ref{fig3} it is sufficient to  approximate
$C_1^{(0)}(q)$  by 
\bea\label{askjsjsuy}
C_1^{(0)}(q)\approx -{2\pi q\tan(\pi g q)\over 1-g}+
C_0^{(1)}(q)C_0^{(1)}(1-q)+
C_0^{(1)}(-q)C_0^{(1)}(1+q)\ .
\eea
The numerical error due to this  approximation is considerably smaller
than   error bars of our  MC data.


\begin{thebibliography}{99}

\bibitem{Weiss}
U. Weiss, Quantum Dissipative Systems, Series in Modern 
Condensed Matter Physics, Vol. 2, second edition 
(World Scientific, Singapore) 1998

\bibitem{Tinkham}
M. Tinkham, Introduction to Superconductivity,
Second Edition (Dover Publ., INC)  2004

\bibitem{Zilberman}
Yu.M. Ivanchenko and L.A. Zil'berman, 
ZhETF Pis. Red. {\bf 8}, 189 (1968) [JETP Lett. {\bf 8},
113 (1968)]

\bibitem{ZilbermanA}
Yu.M. Ivanchenko and L.A. Zil'berman, 
Zh. Eksp. Teor. Fiz. {\bf 55}, 2395 (1968)

\bibitem{Yagi}
R. Yagi, S. Kabayashi and Y. Ootuka,
Jour. of Phys. Soc. of Japan, {\bf 66}, 3722 (1997)


\bibitem{Sonin}

J. Penttil${\rm {\ddot a}}$, ${\rm {\ddot U}}$. Parts, P.J. Hakonen, 
M.A. Paalanen and E.B. Sonin,
Phys. Rev. Lett.,  {\bf 82},  1004 (1999)

\bibitem{Zaikin}

G. Sch${\rm {\ddot o}}$n and A.D. Zaikin,
Phys. Rep. ${\bf 5\&6}$ 237 (1990)


\bibitem{Leggett}

A.O. Caldeira and A.J. Leggett, Phys. Rev. Lett. {\bf 49}, 211 (1981)


\bibitem{Schmid}

A. Schmid, Phys. Rev. Lett. {\bf 51}, 1506 (1983)

\bibitem{Bulgadaev}
S.A. Bulgadaev, JETP Lett. {\bf 39}, 319 (1984)

\bibitem{Guinea}

F. Guinea, V. Hakim and A. Muramatsu,
Phys. Rev. Lett. {\bf 54}, 263 (1985)

\bibitem{Fisher}
M.P.A. Fisher and W. Zwerger,
Phys. Rev. {\bf B32}, 6190 (1985);

\bibitem{Zam}
S. Ghoshal and A.B. Zamolodchikov, Int. J. Mod. Phys. {\bf A9} 3841 (1994)


\bibitem{Warner}
P. Fendley, H. Saleur and N.P. Warner,
Nucl. Phys. {\bf B430}, 577 (1994)

\bibitem{Saleur}
P. Fendley, A.W.W. Ludwig and H. Saleur,
Phys. Rev. Lett. {\bf 74}, 3005 (1995);
P. Fendley, A.W.W. Ludwig and H. Saleur,
Phys. Rev. {\bf B52}, 8934 (1995);



\bibitem{BLZ}
V.V. Bazhanov, S.L. Lukyanov and A.B. Zamolodchikov,
Nucl. Phys. {\bf B549} [FS], 529 (1999);
Jour. of Stat. Phys.  {\bf 102}, 567 (2001)



 
\bibitem{FLZZ}
V.A. Fateev, S.L. Lukyanov, Al.B. Zamolodchikov and A.B. Zamolodchikov,
Phys. Lett. {\bf B406}, 83 (1997)


\bibitem{Herrero02} 
C. P. Herrero and A. D. Zaikin, Phys. Rev.  {\bf B65}, 104516 (2002)

\bibitem{Kimura}
N. Kimura and T. Kato, 
Phys. Rev. {\bf B69}, 012504 (2004)

\bibitem{WT}
P. Werner and M. Troyer,
Phys. Rev. Lett. {\bf 95}, 060201 (2005)

\bibitem{Swendsen87} R. H. Swendsen and J.-S. Wang, 
Phys. Rev. Lett. \textbf{58}, 86 (1987)

\bibitem{Wolff89} U. Wolff, Phys. Rev. Lett. {\bf 62}, 361 (1989)

\bibitem{Luijten95} 
E. Luijten and H. W. K. Bl\"ote, Int. J. Mod. Phys.  {\bf C6}, 359 (1995)

\bibitem{Kane}
K. Moon, H. Yi, C.L. Kane, S.M. Girvin and M.P.A. Fischer,
Phys. Rev. Lett. {\bf 71}, 4381 (1993)

                                                                                  
\bibitem{Klebanov}
C.G. Callan and  I.R. Klebanov,
Phys. Rev. Lett. {\bf 72}, 1968  (1994)
                                                                                  
\bibitem{Maldacena}
C.G. Callan, I.R. Klebanov, A.W.W. Ludwig and  J.M. Maldacena,
Nucl. Phys. {\bf B422}, 417 (1994)
 
\bibitem{Polchinski}
J. Polchinski and  L. Thorlacius,
Phys. Rev. {\bf D50}, 622 (1994)
                                                                                  
\bibitem{Thorlacius}
K.R. Kristjansson and  L. Thorlacius,
JHEP {\bf 0501}, 047 (2005)
                                                                                  
\bibitem{Sen}
A. Sen,
JHEP {\bf 0204}, 048 (2002)

\bibitem
{Calan} C.G. Callan and L. Thorlacius, Nucl. Phys. {\bf B329},
117 (1990)


\bibitem{BLZ2}
V.V. Bazhanov, S.L. Lukyanov and A.B. Zamolodchikov,
Comm. Math. Phys. {\bf 190}, 247 (1997)

\bibitem{Lesage}
F. Lesage and H. Saleur,
Nucl. Phys.  {\bf B546}, 585 (1999)

\bibitem{Lewellen}
J.L. Cardy and D.C. Lewellen,
Phys. Lett. {\bf B259}, 274 (1991)

\bibitem{Lin}

H.H. Lin and M.P.A. Fisher,
Phys. Rev. {\bf B54}, 10593  (1996)


\bibitem{WG}
U. Weiss and H. Grabert,
Phys. Lett. {\bf A108}, 63 (1985)


\bibitem{Refael03}

G. Refael, E. Demler, Y. Oreg, and D. S. Fisher,
Phys. Rev. {\bf B68}, 214515 (2003)

\bibitem{WRT}

P. Werner, G. Refael and M. Troyer,
J. Stat. Mech. P12003 (2005) 

\bibitem{Tewari06}

S. Tewari, J. Toner and S. Chakravarty,
Phys. Rev. {\bf B73}, 064503 (2006)

\bibitem{Goswami06}

P. Goswami and S. Chakravarty,
Phys. Rev. {\bf B73}, 094516 (2006)

\bibitem{Refael07}

G. Refael, E. Demler, Y. Oreg and D. S. Fisher
Phys. Rev. {\bf B75}, 014522 (2007)

\bibitem{ALPS} M. Troyer {\it et al.}, Lecture Notes in Computer Science {\bf 1505}, 191 (1998); J. Phys. Soc. Jpn. Suppl. {\bf 74}, 30 (2005).

\bibitem{ALZ}
                                                                                 
Al.B. Zamolodchikov, Nucl. Phys. {\bf B348}, 619 (1991)

\end{thebibliography}
\end{document}